\documentclass{article}
\usepackage[utf8]{inputenc}
\usepackage[a4paper,margin=1.3in]{geometry}
\usepackage{amsmath,dsfont}
\usepackage{amsthm,subcaption,graphicx}
\usepackage{amsfonts,amssymb}
\usepackage[font=small,labelfont=bf]{caption}
\usepackage{authblk}
\usepackage[authoryear,round]{natbib}
\usepackage{fancyvrb}
\DefineVerbatimEnvironment{Code}{Verbatim}{}
\DefineVerbatimEnvironment{CodeInput}{Verbatim}{fontshape=sl}
\DefineVerbatimEnvironment{CodeOutput}{Verbatim}{}
\newenvironment{CodeChunk}{}{}
\let\proglang=\textsf
\newcommand{\code}[1]{{\normalfont\ttfamily#1}}
\newcommand{\class}[1]{`\code{#1}'}
\newcommand{\fct}[1]{\code{#1()}}
\newcommand{\pkg}[1]{{\fontseries{m}\fontseries{b}\selectfont #1}}

\usepackage{amsmath,dsfont}
\usepackage{amsthm,subcaption,graphicx}
\usepackage{amsfonts,amssymb}
\usepackage{dsfont}
\usepackage{subcaption}
\usepackage{enumitem}
\usepackage{blindtext}
\usepackage{url}
\usepackage{hyperref,xcolor}

\newcommand{\Var}{\mathrm{Var}}
\newcommand{\argmin}[1]{\underset{#1}{\mathrm{argmin}}\ }

\newcommand{\bs}[1]{\boldsymbol{#1}}

\newcommand{\dd}{\mathrm{d}}

\definecolor{oranje}{cmyk}{0,0.50,0.84,0}

\newcommand{\diag}{\mathrm{diag}}

\newtheorem*{remark}{Remark}

\title{\pkg{ecpc}: An \proglang{R}-package for generic co-data models for high-dimensional prediction}
\author[1]{Mirrelijn M. van Nee\thanks{ 
The first author is supported by ZonMw TOP grant COMPUTE CANCER (40-
00812-98-16012).}}
\author[2]{Lodewyk F.A. Wessels}
\author[1,3]{Mark A. van de Wiel}
\affil[1]{{\small{Epidemiology and Data Science, Amsterdam Public Health research institute, Amsterdam University Medical Centers, The Netherlands}}}
\affil[2]{{\small{Molecular Carcinogenesis, Oncode Institute and Netherlands Cancer Institute}}}
\affil[3]{{\small{MRC Biostatistics Unit, Cambridge University, UK}}}
\date{}

\begin{document}

\maketitle

\begin{abstract}
High-dimensional prediction considers data with more variables than samples. 
Generic research goals are to find the best predictor or to select variables. 
Results may be improved by exploiting prior information in the form of co-data, providing complementary data not on the samples, but on the variables. 
We consider adaptive ridge penalised generalised linear and Cox models, in which the variable specific ridge penalties are adapted to the co-data to give a priori more weight to more important variables.
The \proglang{R}-package \pkg{ecpc} originally accommodated various and possibly multiple co-data sources, including categorical co-data, i.e. groups of variables, and continuous co-data.
Continuous co-data, however, was handled by adaptive discretisation, potentially inefficiently modelling and losing information.
Here, we present an extension to the method and software for generic co-data models, particularly for continuous co-data. 
At the basis lies a classical linear regression model, regressing prior variance weights on the co-data.
Co-data variables are then estimated with empirical Bayes moment estimation.
After placing the estimation procedure in the classical regression framework, extension to generalised additive and shape constrained co-data models is straightforward.
Besides, we show how ridge penalties may be transformed to elastic net penalties with the \proglang{R}-package \pkg{squeezy}.
In simulation studies we first compare various co-data models for continuous co-data from the extension to the original method. Secondly, we compare variable selection performance to other variable selection methods.  
Moreover, we demonstrate use of the package in several examples throughout the paper.
\end{abstract}



\section{Introduction}
Generalised linear models (GLMs) \citep{mccullagh1989generalized} are the cornerstone of many statistical models for prediction and variable selection purposes, modelling the relation between outcome data and observed data. 
When observed data are high-dimensional, with the number of variables far exceeding the number of samples, these models may be penalised to account for the high-dimensionality. 
Well known examples include the ridge \citep{hoerl1970ridge}, lasso \citep{tibshirani1996regression} and elastic net penalty \citep{zou2005regularization}.
One of the main assumptions underlying generalised linear models is that all variables are exchangeable. In many high-dimensional settings, however, this assumption is questionable \citep{ignatiadis2020sigma}. 
For example, in cancer genomics, variables may be grouped according to some biological function. Variables within these groups may have a similar effect, while variables from different groups have a different effect. Hence, variables are exchangeable within groups, but not between groups.
represent gene expression of genes that have similar effect within groups representing some biological function but different effect between those groups.
To alleviate the exchangeability assumption, shared information may be modelled explicitly in the prior distribution of the variables, e.g. by introducing shared group penalties, penalising more important groups of variables relatively less (as done by \cite{wiel2016better}). 
The shared prior information may be represented in data matrices, called co-data, to distinguish the main, observed data with information on the samples from the complementary data with information on the variables.
When the co-data are related to the effect sizes of variables, these data may be exploited to improve prediction and variable selection in high-dimensional data settings.

Various \proglang{R}-packages accommodate approaches to incorporate some form of co-data.
Early methods such as \pkg{grplasso} \citep{Meier2008} and \pkg{gglasso} \citep{yang2015fast} allow for categorical, or grouped, co-data, by using group lasso penalties. As these penalties are governed by one overall penalty parameter, these types of penalties may be not flexible enough to model the relation between the effect sizes and grouped co-data.
To increase this flexibility, other methods were developed that estimate multiple, group-specific penalty (or prior) parameters, using efficient empirical Bayes approaches.
Examples include \pkg{GRridge} \citep{wiel2016better} for group-adaptive ridge penalties (normal priors), \pkg{graper} \citep{velten2018adaptive} for group-adaptive spike-and-slab priors and \pkg{gren} \citep{munch2018adaptive} for group-adaptive elastic net priors.
The method \pkg{ecpc} \citep{van2021flexible} presents a flexible empirical Bayes approach to extend the use of grouped co-data to various other (and potentially multiple) co-data types, such as hierarchical groups and continuous co-data, for multi-group adaptive ridge penalties. 
For continuous co-data, however, the normal prior variances corresponding to the ridge penalties are not modelled as a function of the continuous co-data variable, but rather as a function of groups of variables corresponding to the adaptively discretised co-data variable.
When the relation between the prior variance and continuous co-data is non-constant and/or ``simple'', e.g. linear, the adaptive discretisation may lead to a loss of information and/or inefficiently model the relation.
The package \pkg{fwelnet} \citep{tay2020feature} develops feature-weighted elastic net for continuous co-data specifically (there called ``features of features''). Regression coefficients are estimated jointly with co-data variable weights, modelling the variable-specific elastic net penalties by a normalised, exponential function of the co-data.
For categorical co-data, \pkg{fwelnet} boils down to an elastic net penalty on the group level \citep{tay2020feature}, governed by one overall penalty parameter. Hence, it may lack flexibility when compared to empirical Bayes methods estimating multiple penalties.
The package \pkg{squeezy} \citep{vanNee2021fast} presents fast approximate marginal likelihood estimates for group-adaptive elastic net penalties, but is available for grouped co-data only. 

Here, we present an extension of the \proglang{R}-package \pkg{ecpc} to generic co-data models, in particular for continuous co-data.
First, we show how a classical linear regression model may be used to regress the (unknown) variable-specific normal prior variances on the co-data.
The co-data variable weights are estimated with an empirical Bayes moment estimator, slightly modified from \cite{van2021flexible}.
Then, we present how the estimation procedure may be extended straightforwardly to model the relation between the prior variances and co-data by generalised additive models \citep{hastie1986generalized} for modelling non-linear functions and by shape constrained additive models \citep{pya2015shape} for shape constrained functions, e.g. positive and monotonically increasing functions.
Besides, we use ideas from \cite{vanNee2021fast} to transform the adaptive ridge penalties to elastic net penalties using the package \pkg{squeezy}.
Either this approach or the previously implemented posterior selection approaches \citep{van2021flexible} may be used for variable selection.

\subsection{Getting started}
\begin{figure}
    \centering
    \includegraphics[angle=90,width=0.9\textwidth]{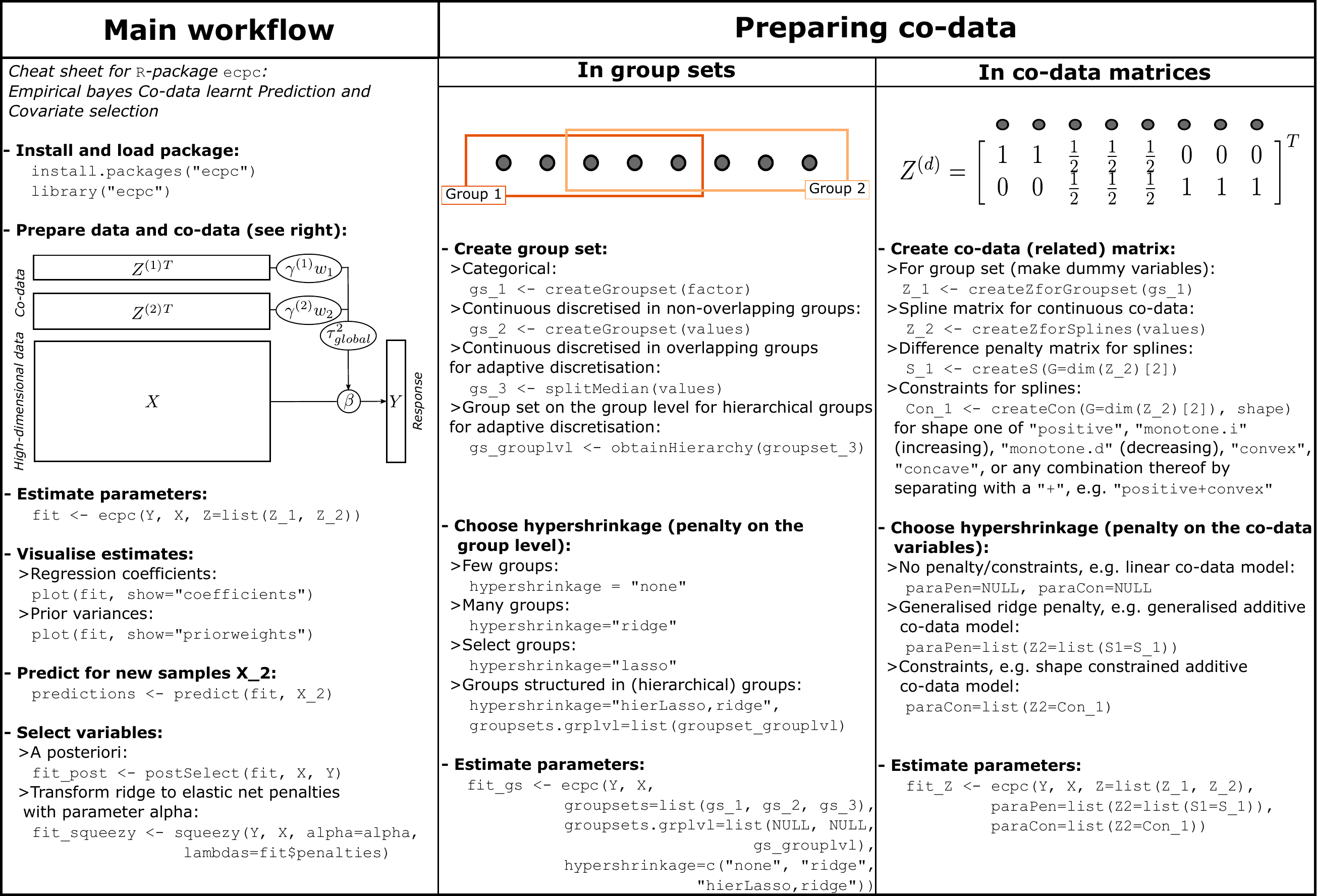}
    \caption{Cheat sheet for the main functions and work flow of the \proglang{R}-package \pkg{ecpc}, available as pdf-file on \url{https://github.com/Mirrelijn/ecpc}.}
    \label{fig:workflow}
\end{figure}

The goal of this paper is to provide a stand-alone introduction to the package \pkg{ecpc} and to provide examples of its use.
To get started, one may install the \proglang{R}-package \pkg{ecpc} from CRAN and load it by running:
\begin{CodeChunk}
\begin{CodeInput}
R> install.packages("ecpc")
R> library("ecpc")
\end{CodeInput}
\end{CodeChunk}
The main function in the package is the eponymous function \fct{ecpc}, which fits a ridge penalised generalised linear model by estimating the co-data variable weights and regression coefficients subsequently. The function outputs an object of the S3-class \class{ecpc}, for which the methods \fct{summary}, \fct{print}, \fct{plot}, \fct{predict} and \fct{coef} have been implemented.
See the index in \code{?{}"ecpc-package"} for a list of all functions, including functions for preparing and visualising co-data, or see Figure \ref{fig:workflow} for a cheat sheet of the main functions and workflow of the package. 

The remainder of this paper is organised as follows: Section \ref{sec:method} first presents the model and the co-data models for a linear, generalised additive and shape-constrained co-data model.
All types of co-data models are accompanied with short examples of how to use the package.
Next, it is shown how ridge penalties may be transformed to elastic net penalties, again accompanied with a short toy example.  
Section \ref{sec:simulation} first compares various ways of modelling a continuous co-data variable with the originally proposed adaptive discretisation in a simulation study. Secondly, the method is compared to other methods in terms of variable selection.
Section \ref{sec:LNManalysis} then demonstrates use of the software on an application to the classification of lymph node metastasis using high-dimensional RNA expression data.
Section \ref{sec:conclusion} shortly concludes the software.

\section{Method}\label{sec:method}
Consider response data $\bs{Y}\in\mathbb{R}^n$, observed high-dimensional data $X\in\mathbb{R}^{n\times p}$ with $p\gg n$, which contain information on the $n$ samples of $\bs{Y}$, and possibly multiple co-data matrices $Z^{(d)}\in\mathbb{R}^{p\times G_d}$, $d=1,..,D$, which contain prior information on the $p$ variables of $X$. 
Generally, co-data matrices may include continuous or categorical co-data. For categorical co-data, dummy variables should be provided. For categorical co-data with overlapping categories, dummy variables may be weighted accordingly to account for multiplicity (see \cite{van2021flexible}). 

We consider a generalised linear model for the response with canonical link function $g(\cdot)$, parameterised with regression coefficients $\bs{\beta}\in\mathbb{R}^p$. Furthermore, we model the regression coefficients with a normal prior, corresponding to a ridge penalty, in which the prior variance is regressed on the co-data:
\begin{align}\label{eq:model}
\begin{split}
    Y_i | \bs{X}_i,\bs{\beta} &\overset{ind.}{\sim} \pi\left(Y_i | \bs{X}_i,\bs{\beta}\right),\ E_{Y_i|\bs{X}_i,\bs{\beta}}(Y_i)=g^{-1}(\bs{X}_i\bs{\beta}),\ i=1,..,n,\\
    \beta_{k}&\overset{ind.}{\sim} N(0,v_k),\ v_k=\tau_{global}^2\sum_{d=1}^D w_d \bs{Z}_k^{(d)} \bs{\gamma}^{(d)},\ k=1,..,p, 
\end{split}
\end{align}
with $\bs{X}_i$ and $\bs{Z}_k$ the $i^{th}$ and $k^{th}$ row of $X$ and $Z$ respectively, $\bs{\gamma}^{(d)}\in\mathbb{R}^G$ the co-data variable weights for co-data matrix $d$, $\bs{w}$ the co-data matrix weights and $\tau_{global}^2$ a scaling factor which may improve numerical computations in practice. When the data $X$ consist of multiple data modalities, like gene expression data, copy number data and methylation data in genomics, scaling factors specific to the data modalities may be used \citep{boulesteix2017ipf,van2021fast} and estimated with \pkg{ecpc}.

Prior parameters and regression coefficients are estimated with an empirical Bayes approach, following \cite{van2021flexible}. In short, first the global scaling parameter $\tau^2_{global}$ is estimated, then the co-data variable weights $\bs{\gamma}^{(d)}$ for each co-data matrix $d$ separately and then the co-data weights $\bs{w}$. 
After, given the prior parameter estimates, the regression coefficients are estimated by maximising the penalised likelihood (equivalent to maximising the posterior). 
Multiple co-data matrices may be either provided in different matrices $Z^{(1)},..,Z^{(D)}$ or stacked and provided in one matrix $Z:=[Z^{(1)},..,Z^{(D)}]$. Below, we first consider the case in which we have only one (stacked) co-data matrix $Z$, for which $\bs{w}=1$, and drop notions of $d$. Besides, without loss of generality, we set the scaling parameter $\tau_{global}^2$ to $1$. See the remark in Section \ref{sec:scam} for discussion of the differences between multiple matrices and one stacked matrix. 

Next, we show how the empirical Bayes approach of estimating $\bs{\gamma}$ \citep{van2021flexible} straightforwardly adapts to continuous co-data when the relation is assumed to be linear (as in Equation \eqref{eq:model}).
From thereon, we show how the approach naturally fits into the framework of generalised additive models and shape constrained additive models.

\begin{remark}
The first version of \pkg{ecpc}, as described in \cite{van2021flexible}, only handles (possibly overlapping) groups of variables. 
Co-data is then supplied in a list of group sets, in the argument \code{groupsets}.
Continuous co-data variables may be handled by adaptive discretisation.
The new version discussed in this paper allows for both (undiscretised) continuous and grouped co-data.
In the new version of \pkg{ecpc}, one needs to supply co-data as a list of co-data matrices in the argument \code{Z}.
The function \fct{createZforGroupset} may be used to obtain the co-data matrix with dummy variables corresponding to a group set for grouped co-data.
\end{remark}

\subsection{Details for linear co-data models}\label{sec:continuous}
Consider a linear relation between the prior variances and the co-data:
\begin{align*}
    \bs{v}= Z\bs{\gamma}.
\end{align*}
The co-data variable weights $\bs{\gamma}$ are estimated in \cite{van2021flexible} with moment estimation by equating theoretical moments to empirical moments. For co-data that represent groups of variables, the empirical moments are averaged over all variables in that group, leading to a linear system of $G$ equations and $G$ unknowns. For co-data that do not represent groups of variables, e.g. continuous co-data, we simply form one group per variable, leading to the following linear system of $p$ equations and $G$ unknowns:
\begin{align}\label{eq:gammaest}
        &(C\circ C)Z\bs{\gamma}=\bs{b},
\end{align}
with $\circ$ representing the Hadamard (element-wise) product. 
$C\in\mathbb{R}^{p\times p}$ and $\bs{b}\in\mathbb{R}^{p}$ are derived in \cite{van2021flexible} and given by:
\begin{align*}
        &C=(X^TWX+\tilde{\Omega})^{-1}X^TWX,\nonumber\\
        &\bs{b}=\tilde{\bs{\beta}}.^2-\tilde{\bs{v}},\nonumber\\
        &\tilde{\bs{v}}=\diag((X^TWX+\tilde{\Omega})^{-1}X^TWX(X^TWX+\tilde{\Omega})^{-1}),\nonumber
\end{align*}
with $\tilde{\bs{\beta}}$ the maximum penalised likelihood estimate given an initial $\tilde{\tau}^2_{global}$ and corresponding constant diagonal ridge penalty matrix $\tilde{\Omega}$, with $W$ a diagonal weight matrix used in the iterative weighted least squares algorithm to fit $\tilde{\bs{\beta}}$, and with $\tilde{\bs{v}}$ an estimate for the variance of $\tilde{\bs{\beta}}$ with respect to the response data $Y$.

Solving the linear system leads to the following least squares estimate for $\bs{\gamma}$. As the prior variance has to be positive, the resulting prior variance estimate $Z\bs{\gamma}$ is truncated at $0$:
\begin{align}\label{eq:MoMlinear}
    \hat{\bs{\gamma}} &= \argmin{\bs{\gamma}} ||(C\circ C)Z\bs{\gamma}-\bs{b}||^2_2, \qquad \hat{\bs{v}}=(Z\hat{\bs{\gamma}})_+.
\end{align}
In generalised linear models it is common to use a log-link for the response to enforce positivity, resulting in positive, multiplicative effects. Note that here, however, Equation \eqref{eq:gammaest} is the result of equating theoretical to empirical moments. Replacing $\bs{b}$ by $\log(\bs{b})$ would violate the moment equalities. 
Also, if we would enforce positivity instead by, for example, substituting $Z\bs{\gamma}$ directly by $\bs{v}=\exp(Z\bs{\gamma}')$, the moment equations would not be linear anymore in $\bs{\gamma}$, nor multiplicative, e.g. as $(C\circ C)\exp(Z\bs{\gamma})\neq \exp((C\circ C)Z\bs{\gamma})=\prod_{g=1}^G\exp((C\circ C)\bs{Z}_g\gamma_g)$, with $\bs{Z}_g$ the $g^{th}$ co-data variable.
Hence, the advantage of simply post-hoc truncating $Z\hat{\bs{\gamma}}$ is that the system of equations in Equation \eqref{eq:MoMlinear} is easily solved.
Section \ref{sec:scam} discusses shape constrained co-data models which may be used to enforce positivity by including it as a constraint.  

\subsubsection{Interpretation}
The interpretation of the co-data weights $\bs{\gamma}$ (scaled by $\tau_{global}^2$) is similar as in regular linear regression: when co-data variable $Z_g$ increases with one unit, while the other co-data variables are kept fixed, then the prior variance increases with $\gamma_g$ ($\gamma_g\tau_{global}^2$). Consequently, when the prior variance for the effect $\beta_k$ of some variable $X_k$ increases with $\gamma_g$ ($\gamma_g\tau_{global}^2$), then the a priori expected squared effect size, $E(\beta_k^2)=v_k$, increases with $\gamma_g$ ($\gamma_g\tau_{global}^2$). In other words, when we would compare the effect of two variables $X_k$ and $X_l$ with the same co-data values, except for one co-data variable which is one unit higher for $X_k$ than for $X_l$, then we would a priori expect $\beta_k^2$ to be on average $\gamma_g$ ($\gamma_g\tau_{global}^2$) larger than $\beta_l^2$.

\subsubsection{Short example in ecpc}
For this short example and the ones below, simulate some linear response data:
\begin{CodeChunk}
\begin{CodeInput}
R> set.seed(1)
R> p <- 300 #number of covariates
R> n <- 100 #sample size training data set
R> n2 <-100 #sample size test data set
R> beta <- rnorm(p, mean=0, sd=0.1) #simulate effects
R> X <- matrix(rnorm(n*p, mean=0, sd=1), n, p) #simulate observed training data
R> Y <- rnorm(n, mean = X%*%beta, sd=1) #simulate response training data
R> X2 <- matrix(rnorm(n2*p, mean=0, sd=1), n, p) #simulate observed test data
R> Y2 <- rnorm(n2, mean = X2%*%beta, sd=1) #simulate response test data
\end{CodeInput}
\end{CodeChunk}
As co-data, suppose that we have two co-data variables; one informative co-data variable containing the true absolute effect sizes and one non-informative co-data variable containing random normally distributed values:
\begin{CodeChunk}
\begin{CodeInput}
R> Z1 <- abs(beta) #informative co-data
R> Z2 <- rnorm(p, mean=0, sd=1) #random, non-informative co-data
R> Z <- cbind(Z1, Z2) #(px2)-dimensional co-data matrix 
\end{CodeInput}
\end{CodeChunk}
Then we fit the linear co-data model and test the fit on the test data. 
Besides, we set \code{postselection=FALSE} to only estimate the dense model, without selecting variables a posteriori (see Section \ref{sec:transform}):
\begin{CodeChunk}
\begin{CodeInput}
R> fit <- ecpc(Y, X, Z=list(Z), X2=X2, Y2=Y2, postselection=FALSE)
\end{CodeInput}
\begin{CodeOutput}
[1] "Estimate global tau^2 (equiv. global ridge penalty lambda)"
[1] "Estimate co-data weights and (if included) hyperpenalties with mgcv"
[1] "Estimate regression coefficients"
\end{CodeOutput}
\end{CodeChunk}
Note that the co-data matrix is provided in a list, as it is also possible to provide a list of multiple co-data matrices. This will be used to explicitly distinguish linear co-data variables from smooth or constrained ones, as exemplified below.
The performance of the fit on the test data may be given for both the co-data learnt model fit with \fct{ecpc} and for the co-data agnostic model fit with one global ridge penalty:
\begin{CodeChunk}
\begin{CodeInput}
R> fit$MSEecpc 
\end{CodeInput}
\begin{CodeOutput}
[1] 2.521757
\end{CodeOutput}
\begin{CodeInput}
R> fit$MSEridge
\end{CodeInput}
\begin{CodeOutput}
[1] 2.889294
\end{CodeOutput}
\end{CodeChunk}
A (summary of) the fitted prior parameters, prior variances and regression coefficients can be retrieved by the methods \fct{print} and \fct{summary}:
\begin{CodeChunk}
\begin{CodeInput}
R> print(fit) 
\end{CodeInput}
\begin{CodeOutput}
ecpc fit

Estimated co-data variable weights: 
0.2125547 -0.001632461  

Estimated co-data weights: 
1
\end{CodeOutput}
\begin{CodeInput}
R> summary(fit) 
\end{CodeInput}
\begin{CodeOutput}
Summary estimated prior variances: 
     Min.   1st Qu.    Median      Mean   3rd Qu.      Max. 
0.0000000 0.0000000 0.0008287 0.0068645 0.0106837 0.0471067 

Summary estimated regression coefficients: 
     Min.   1st Qu.    Median      Mean   3rd Qu.      Max. 
-0.251409  0.000000  0.000000  0.002368  0.002004  0.269845 

Estimated intercept:  0.07294278  
\end{CodeOutput}
\end{CodeChunk}
Alternatively, the \fct{plot} method provides a graph of the regression coefficients and prior variances. If the \proglang{R}-packages \pkg{ggplot2} \citep{ggplot2} and \pkg{ggpubr} \citep{kassambara2020package} are installed, the output looks as shown in Figure \ref{fig:outputexample1}, else a similar plot will be made with the base \proglang{R} \fct{plot} function.
\begin{CodeChunk}
\begin{CodeInput}
R> plot(fit, show="coefficients")
R> plot(fit, show="priorweights", Z=list(Z))
\end{CodeInput}
\begin{figure}[h]
    \centering
    \includegraphics[width=0.8\textwidth]{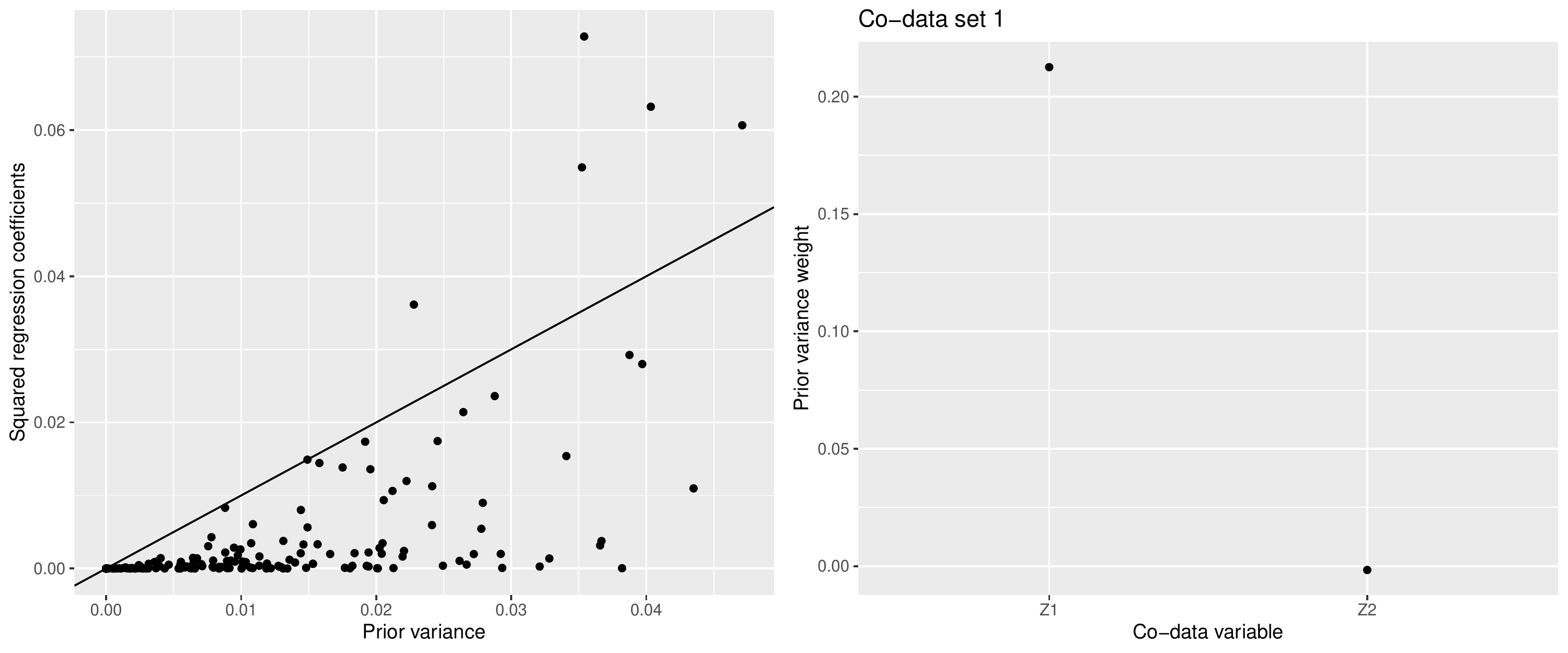}
    \caption{Example output for \fct{plot} in a linear co-data model for \code{show="coefficients"} (left) and \code{show="priorweights"} (right).}
    \label{fig:outputexample1}
\end{figure}
\end{CodeChunk}
Lastly, the regression coefficients may be re-estimated for different prior parameters. First, the function \fct{penalties} may be used to change some prior parameters and retrieve the corresponding ridge penalties. Then, the method \fct{coef} re-estimates the regression coefficients given these penalties.
For example, if one would alter the global level of regularisation by multiplying $\tau^2_{global}$ by 2:
\begin{CodeChunk} 
\begin{CodeInput}
R> new_penalties <- penalties(fit, tauglobal = fit$tauglobal * 2, Z=list(Z)) 
R> new_coefficients <- coef(fit, penalties=new_penalties, X=X, Y=Y) 
\end{CodeInput} 
\end{CodeChunk}

Note that in general, however, altering the prior parameters by hand will not be needed as these parameters are optimised by the function \fct{ecpc}. 
The functions above, however, may be used to conveniently skip prior parameter estimation when prior parameters are known, e.g. when results have been saved and need to be checked quickly. For example, here we just set all prior parameters to one and only estimate the regression coefficients:
\begin{CodeChunk} 
\begin{CodeInput}
R> new_penalties2 <- penalties(tauglobal = 1, sigmahat = 1, gamma = c(1,1), 
+    w = 1, Z=list(Z)) 
R> new_coefficients2 <- coef.ecpc(penalties=new_penalties2, X=X, Y=Y) 
\end{CodeInput} 
\end{CodeChunk}

\subsection{Details for generalised additive co-data models}\label{sec:gam}
Generalised additive models (GAMs), originally proposed in \cite{hastie1986generalized}, have been widely applied to model non-linear relations. Applied here, we assume that the relation between the prior variance and co-data may be modeled by a sum of smooth functions, $s_1(\cdot),..,s_G(\cdot)$, of the co-data variables:
\begin{align*}
    \bs{v} = \sum_{g=1}^G s_g(\bs{Z}_g).
\end{align*}
In practice, the smooth functions are estimated by using a basis expansion to recast the problem into a linear model (as originally proposed by, for example, \cite{wahba1980spline}). So, for a basis expansion consisting of $J_g$ basis functions $\phi_{g,j}(\cdot)$, $j=1,..,J_g$, for co-data variable $\bs{Z}_g$:
\begin{align*}
    s_g(\bs{Z}_g) &= \sum_{j=1}^{J_g} \phi_{g,j}(\bs{Z}_g)\gamma_{g,j} = \Phi_j\bs{\gamma}_g,\ \qquad \bs{v} = \sum_{g=1}^G \Phi_j\bs{\gamma}_g = Z_{GAM}\bs{\gamma}_{GAM},
\end{align*}
with $\Phi_g\in\mathbb{R}^{p\times J_g}$ the matrix of co-data variable vector $\bs{Z}_g\in\mathbb{R}^p$ evaluated in all $J_g$ basis functions, $Z_{GAM}=[\Phi_1, .., \Phi_G]$ and $\bs{\gamma}_{GAM}=(\bs{\gamma}_{1}^T,..,\bs{\gamma}_{G}^T)^T$.

The type and number of basis functions should in general be chosen such that they are flexible enough to approximate the underlying function well. 
To avoid overfitting for too many basis functions, the coefficients may be estimated by optimising the likelihood penalised by a smoothing penalty. 
While our software allows the user to supply any basis expansion, we focus here on the popular p-splines (see \cite{eilers2021practical} for an introduction).
This approach combines flexible spline basis functions with a quadratic smoothing penalty on the differences of the spline coefficients.
So, the smoothing penalty is of the form $\bs{\gamma}_{GAM}^T\left(\sum_{g}\lambda_g S_g\right) \bs{\gamma}_{GAM}$, where the difference penalty matrix $S_g$ smooths the non-linear function of the co-data variable $\bs{Z}_g$ and where $\lambda_g$ is the corresponding smoothing penalty parameter.
Hence, the least squares estimate for the linear co-data model in Equation \eqref{eq:MoMlinear} is extended to the following estimate for the GAM coefficients in a non-linear co-data model:
\begin{align}\label{eq:estGAM}
    \hat{\bs{\gamma}}_{GAM} &= \argmin{\bs{\gamma}} \left\{ ||(C\circ C)Z_{GAM}\bs{\gamma}-\bs{b}||^2_2 + \sum_{g=1}^G \lambda_g \bs{\gamma}^TS_g\bs{\gamma}\right\}, \qquad \hat{\bs{v}}=(Z_{GAM}\hat{\bs{\gamma}}_{GAM})_+.
\end{align}
This least-squares equation is of a form also known as penalised signal regression \citep{marx1999generalized} and can be solved by the function \fct{gam} (or \fct{bam} for big data) of the \proglang{R}-package \pkg{mgcv}, for example. This function also provides fast and stable estimation of the penalties $\lambda_g$ \citep{wood2011fast}. 
Alternatively, when only one smoothing penalty matrix is provided, the smoothing penalty may be estimated by using random splits as proposed in \cite{van2021flexible}.

\begin{remark}
Note that grouped co-data may be coded as group sets or as dummies in a co-data matrix Z. 
The former option, however, does not allow for a generalised ridge penalty, but for other penalties including the ordinary ridge and (hierarchical) lasso penalty.
\end{remark}

\subsubsection{Short example in ecpc}
We continue with the simulated data from above. First, we use the helper functions \\ \fct{createZforSplines} and \fct{createS} to create spline basis matrices and corresponding smoothing penalty matrices respectively. The degree of the spline functions and order of the penalty matrices are set to $3$ and $2$ by default, respectively. We set the number of splines to $20$ for the first co-data variable and to $30$ for the second in this example.
\begin{CodeChunk} 
\begin{CodeInput}
R> Z1.s <- createZforSplines(values=Z1, G=20, bdeg=3) 
R> S1.Z1 <- createS(orderPen=2, G=20) 
R> Z2.s <- createZforSplines(values=Z2, G=30, bdeg=3) 
R> S1.Z2 <- createS(orderPen=2, G=30)
\end{CodeInput} 
\end{CodeChunk}
Before we fit the model, we first concatenate the two co-data matrices in a list. The variables of this list are always renamed such that the $i^{th}$ element is named \code{Zi}. 
The smoothing penalty matrices should be given in a separate argument \code{paraPen}, similar to the eponymous argument in \fct{gam}. Each element in this argument's list should match one of the names \code{Zi}, for which the corresponding smoothing matrix is given in \code{S1} (and optionally \code{S2}, \code{S3}, et cetera for multiple smoothing matrices for one co-data matrix).
\begin{CodeChunk} 
\begin{CodeInput}
R> Z.all <- list(Z1=Z1.s, Z2=Z2.s)
R> paraPen.all <- list(Z1=list(S1=S1.Z1), Z2=list(S1=S1.Z2))
\end{CodeInput} 
\end{CodeChunk}
Then we fit the model and test it on the test data as follows. Note that an intercept is included by default:
\begin{CodeChunk} 
\begin{CodeInput}
R> fit.gam <- ecpc(Y, X, Z = Z.all, paraPen = paraPen.all, 
+    intrcpt.bam=TRUE, X2=X2, Y2=Y2, postselection=FALSE)
\end{CodeInput}
\begin{CodeOutput}
[1] "Estimate global tau^2 (equiv. global ridge penalty lambda)"
[1] "Estimate co-data weights and (if included) hyperpenalties with mgcv"
[1] "Estimate regression coefficients"
\end{CodeOutput}
\begin{CodeInput}
R> fit.gam$MSEecpc
\end{CodeInput}

\begin{CodeOutput}
[1] 2.472784
\end{CodeOutput}
% \begin{CodeInput}
% R> fit.gam$gamma*fit.gam$tauglobal #fitted co-data spline coefficients
% R> fit.gam$gamma0*fit.gam$tauglobal #fitted co-data intercept 
% \end{CodeInput} 
\end{CodeChunk}
The non-linear relation between the prior variance and each co-data source may again be plotted with the \fct{plot} method, either with the co-data spline variables on the x-axis, or the continuous co-data values on the x-axis. 
The corresponding output is shown in Figure \ref{fig:outputexample2}:
\begin{CodeChunk} 
\begin{CodeInput}
R> plot(fit.gam, show="priorweights", Z=Z.all) 
R> values <- list(Z1, Z2)
R> plot(fit.gam, show="priorweights", Z=Z.all, values = values) 
\end{CodeInput} 
\begin{figure}[h]
    \centering
    \includegraphics[width=\textwidth]{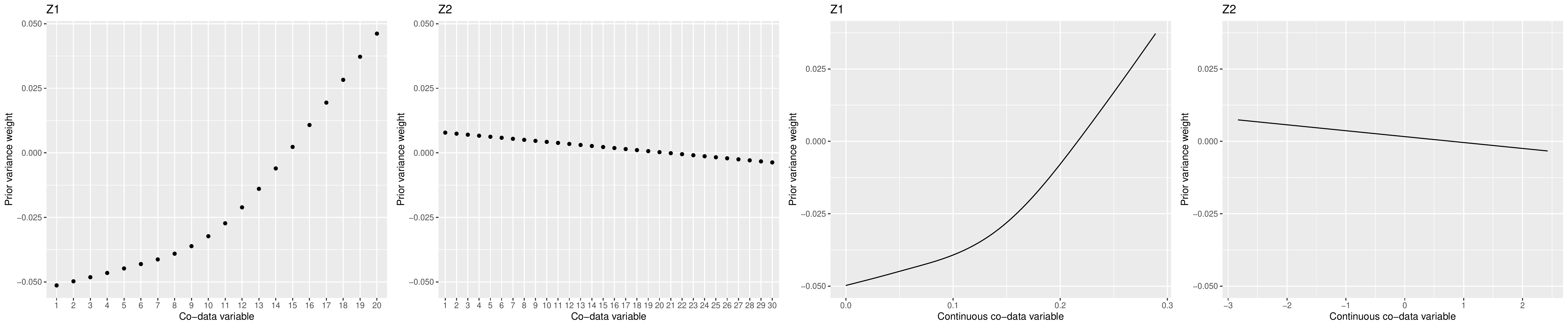}
    \caption{Example output for \fct{plot} in a generalised additive co-data model with the co-data variables on the $x$-axis (left two plots) or continuous values on the $x$-axis (right two plots).}
    \label{fig:outputexample2}
\end{figure}
\end{CodeChunk}
Alternatively, one may plot the non-linear relation directly.
The spline variable coefficients are given in \code{fit\$gamma} for both co-data matrices and have an attribute \code{codataSource} to indicate for each coefficient to which co-data matrix it belongs. 
The non-linear relation of one co-data matrix is then plot as follows (output not shown):
\begin{CodeChunk} 
\begin{CodeInput}
R> codataNO <- attributes(fit.gam$gamma)$codataSource
R> i <- 2 #1 for informative, 2 for non-informative
R> sk <- as.vector(Z.all[[i]]%*%fit.gam$gamma[codataNO==i])*fit.gam$tauglobal
R> par(mfrow=c(1,1))
R> plot(Z[,i],sk)
\end{CodeInput}
\end{CodeChunk}

\subsection{Details for shape-constrained co-data models}\label{sec:scam}
Prior assumptions on the shape of the relation between the prior variance and co-data, such as monotonicity or convexity, may be imposed by constrained optimisation of spline coefficients \citep{pya2015shape}. 
\cite{pya2015shape} develop shape-constrained p-splines to handle difficulties in optimising multiple smoothing penalties due to discontinuous gradients.
Their \proglang{R}-package \pkg{scam}, however, cannot be readily used for signal regression, which differs from regular regression in that the spline basis matrix is multiplied by the known matrix $(C\circ C)$. 
Moreover, the smoothing parameter estimates are estimated using a generalised cross-validation (GCV) criterion, which we show below to overfit in the unconstrained case. 
Therefore, we fall back to the simple approach of directly constraining the spline coefficients. 

We use the approach proposed in \cite{van2021flexible} to estimate the smoothing penalties: first we estimate the smoothing penalties $\lambda_g$ separately for each co-data variable $\bs{Z}_g$ using random splits of the data. As this optimisation is in one dimension only, we can use Brent's algorithm from the general purpose optimisation \proglang{R}-package \pkg{optim}, which should be sufficient to handle discontinuous gradients.
Then we estimate the spline coefficients $\bs{\gamma}_g$ for each co-data variable $\bs{Z}_g$ and corresponding spline basis function matrix $\Phi_g$. 
In the general constrained setting, this estimate is given by subjecting the possible solution to Equation \eqref{eq:estGAM} to (in)equality constraints given in matrix $M_{(in)eq,g}$ and vector $\bs{b}_{(in)eq,g}$:
\begin{align}\label{eq:estSCAM}
\begin{split}
    &\left\{\begin{array}{l}
         \hat{\bs{\gamma}}_g = \argmin{\bs{\gamma}} \left\{ ||(C\circ C)\Phi_g\bs{\gamma}-\bs{b}||^2_2 + \lambda_g \bs{\gamma}^TS_g\bs{\gamma}\right\} \\
         \qquad \textrm{s.t.}\ M_{ineq,g}\bs{\gamma}\leq \bs{b}_{ineq,g},\ M_{eq,g}\bs{\gamma}=\bs{b}_{eq,g} 
    \end{array}\right.
\end{split}
\end{align}
Several shapes may be imposed by choosing $M_{ineq}$ and $b_{ineq}$ accordingly \citep{pya2015shape}: i) positivity may be imposed by constraining the spline coefficients to be positive; ii) monotonically increasing (decreasing) may be imposed by constraining the first order differences $\gamma_{i+1}-\gamma_i$ to be positive (negative); iii) convexity (concavity) may be imposed by constraining second order differences $\gamma_{i+2} - 2\gamma_{i+1} + \gamma_i$ to be positive (negative); iv) any combination of the shapes i-iii may be imposed by combining the corresponding constraints.

Then, given the spline coefficient estimates $\hat{\bs{\gamma}}_g$, we combine multiple co-data variables by estimating co-data source weights $\bs{w}=(w_1,..,w_G)^T$ using the same method of moment equation \citep{van2021flexible}. For $Z_w:=\left[\Phi_1\hat{\bs{\gamma}}_1,..,\Phi_1\hat{\bs{\gamma}}_G\right]$:
\begin{align}\label{eq:MoMweights}
    \hat{\bs{w}} &= \argmin{\bs{w}} ||(C\circ C)Z_w\bs{w}-\bs{b}||^2_2, \qquad \hat{\bs{v}}=\left(\sum_{g=1}^G \hat{w}_g\Phi_g\hat{\bs{\gamma}}_g\right)_+.
\end{align}
Note that a similar equation is used when multiple co-data matrices $Z^{(1)},..,Z^{(D)}$ are provided.

\begin{remark}
Multiple co-data matrices $Z^{(1)},..,Z^{(D)}$ may be provided in a list to the function \fct{ecpc}, or stacked and provided in a list of one co-data matrix $Z=[Z^{(1)},..,Z^{(D)}]$. 
When the function \fct{bam} from \pkg{mgcv} is used, multiple smoothing parameters may be used for either representation, and are estimated jointly. After, the co-data variable weights $\bs{\gamma}$ are jointly estimated for all co-data matrices as well. As a result, the co-data weights $\bs{w}$ do not need to be estimated as they are implicitly accounted for in the joint estimate of $\bs{\gamma}$.
In contrast, when the random splitting is used, only one smoothing parameter per co-data matrix may be estimated. Therefore, the co-data matrix weights are estimated to combine multiple co-data matrices.
By default, the function \fct{ecpc} uses \fct{bam} when co-data is provided in co-data matrices and no constraints are provided. This may be changed by setting \code{hypershrinkage="none"} when no penalty for the moment estimates is used or to \code{hypershrinkage="ridge"} when a generalised ridge penalty as in Equation \eqref{eq:estGAM} is used with random splits for estimating the penalty parameter.
When constraints are provided, the function \fct{ecpc} automatically switches to the random splits.
\end{remark}

\subsubsection{Short example in ecpc}
We continue the short example from above for shape-constrained functions.
Say we would like to find a positive and monotonically increasing function for the first co-data variable, and a convex function for the second variable.
We can use the helper function \fct{createCon} to obtain the constraint matrix $M_{ineq}$ and vector $\bs{b}_{ineq}$ in the desired format for argument \code{paraCon}:
\begin{CodeChunk} 
\begin{CodeInput}
R> Con.Z1 <- createCon(G=20, shape="positive+monotone.i") 
R> Con.Z2 <- createCon(G=30, shape="convex") 
R> paraCon <- list(Z1=Con.Z1, Z2=Con.Z2)
\end{CodeInput} 
\end{CodeChunk}
Then we fit the model and plot the estimated shape-constrained functions as follows, with the output shown in Figure:
\begin{CodeChunk} 
\begin{CodeInput}
R> fit.scam <- ecpc(Y, X, Z = Z.all, paraPen = paraPen.all, 
+    paraCon = paraCon, X2=X2, Y2=Y2, postselection=FALSE)
\end{CodeInput}
\begin{CodeOutput}
[1] "Estimate global tau^2 (equiv. global ridge penalty lambda)"
[1] "Co-data matrix 1: estimate hyperlambda for ridge+constraints hypershrinkage"
[1] "Estimate weights of co-data source 1"
[1] "Co-data matrix 2: estimate hyperlambda for ridge+constraints hypershrinkage"
[1] "Estimate weights of co-data source 2"
[1] "Estimate co-data source weights"
[1] "Estimate regression coefficients"
\end{CodeOutput}
\begin{CodeInput}
R> fit.scam$MSEecpc
\end{CodeInput}

\begin{CodeOutput}
[1] 2.490368
\end{CodeOutput}
\begin{CodeInput}
R> plot(fit.scam, show="priorweights", Z=Z.all, values=values)
\end{CodeInput} 
\begin{figure}[h]
    \centering
    \includegraphics[width=0.8\textwidth]{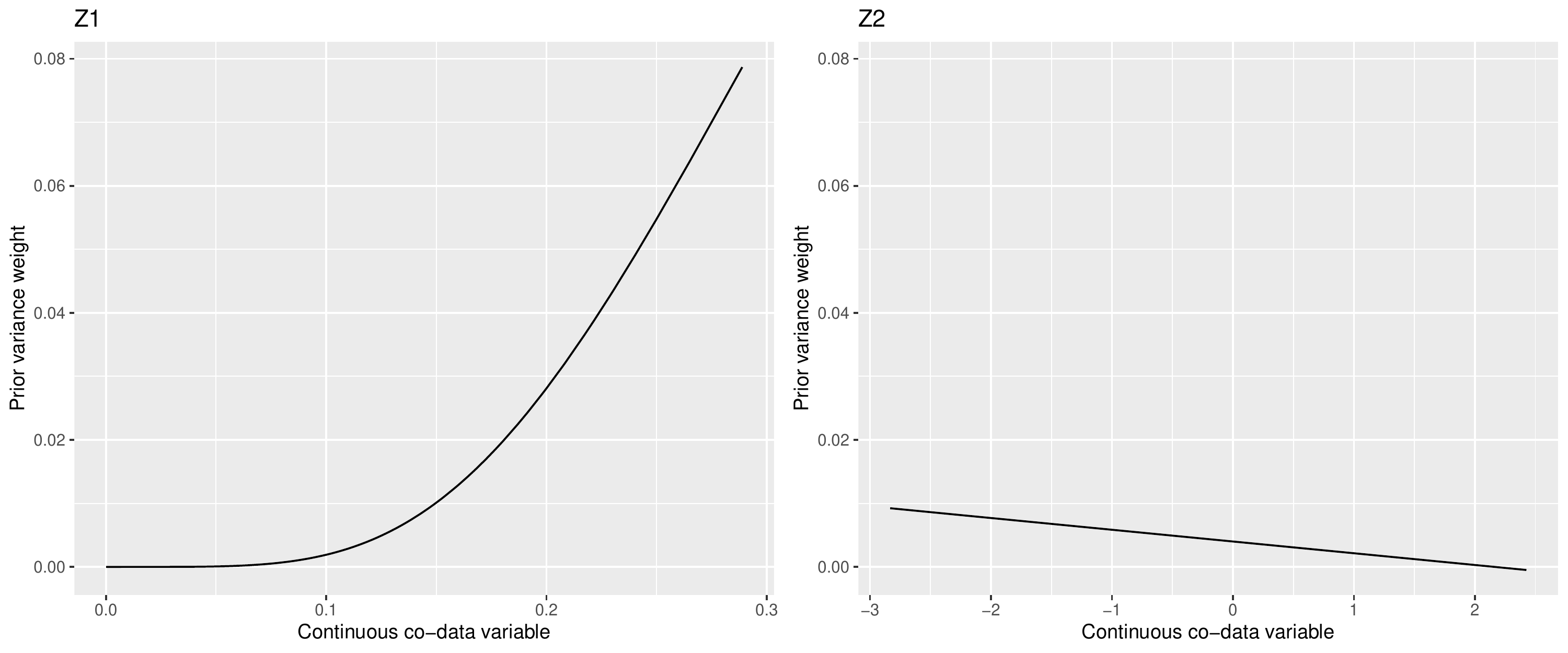}
    \caption{Example output for \fct{plot} in a shape constrained additive co-data model.}
    \label{fig:outputexample3}
\end{figure}
\end{CodeChunk}

Note that an intercept is excluded by default, but that it can easily be included by appending a column of ones to $Z$.


\subsection{Transforming ridge penalties to elastic net penalties}\label{sec:transform}
The first version of \pkg{ecpc} allows for posterior selection \citep{van2021flexible}, exemplified in Section \ref{sec:LNManalysis}. 
Alternatively, the obtained adaptive ridge penalties may be transformed to elastic net penalties for simultaneous estimation and variable selection, as explained here.

In the proposed model in Equation \eqref{eq:model}, the regression coefficients follow a normal prior corresponding to a ridge penalty.
Now, suppose that each $\beta_k$ independently follows some other prior distribution $\pi(\beta_k)$, parameterised by covariate-specific prior parameter $\lambda_k$ and with prior mean $0$ and finite variance $\Var(\beta_k)=Z\bs{\gamma}=h(\lambda_k)$ for some known monotonic variance function $h(\cdot)$:
\begin{align}
    \beta_k\overset{ind.}{\sim}\pi(\beta_k),\ E(\beta_k)=0,\ \Var(\beta_k)=h(\lambda_k)=\bs{Z}_k\bs{\gamma}.
\end{align}
As example we consider the elastic net prior, corresponding to the elastic net penalty, with variable specific elastic net penalty.
Recently, it was shown that when the prior parameters are group-specific, the marginal likelihood -as function of $\lambda_k$- is approximately the same as the marginal likelihood as function of normal prior parameters $\bs{\gamma}$, as the prior distribution of the linear predictor $\bs{\eta}=X\bs{\beta}$ is asymptotically normally distributed \citep{vanNee2021fast}:
\begin{align*}
    \pi(\bs{Y}|X,\bs{\lambda})\approx \pi(\bs{Y}|X,\bs{\gamma})
\end{align*}
This result also holds for priors with variable specific, finite variance \cite{eicker1966multivariate}. 
We may use this result to obtain approximate method of moment equations for other priors.

Denote by $\hat{\bs{\beta}}_R(Y)$ the ridge penalised maximum likelihood estimate as function of the observed response data $\bs{Y}$. The method of moments equations are given by equating the theoretical marginal moments to the empirical moments \citep{van2021flexible}:
\begin{align*}
    E_{\bs{Y}|\bs{\lambda}}(\hat{\beta}_{k,R}^2(\bs{Y})) = \hat{\beta}_{k,R}^2(\bs{Y}),\ \text{for } k=1,..,p.
\end{align*}
Using the normal approximation for the marginal likelihood we obtain:
\begin{align*}
    E_{\bs{Y}|\bs{\lambda}}\left(\hat{\beta}_{k,R}^2(\bs{Y})\right) &= \int_{\bs{Y}} \hat{\beta}_{k,R}^2(\bs{Y}) \pi(\bs{Y}|X,\bs{\lambda}) \dd\bs{Y}\\ 
    &\approx \int_{\bs{Y}}\hat{\beta}_{k,R}^2(\bs{Y}) \pi(\bs{Y}|X,\bs{\gamma})\dd\bs{Y} = E_{\bs{Y}|\bs{\gamma}}\left(\hat{\beta}_{k,R}^2(\bs{Y})\right).
\end{align*}
So we may obtain the ridge estimates $\hat{\bs{\gamma}}$ as above to estimate the covariate specific prior variances $\hat{v}_k=(\bs{Z}_k\hat{\bs{\gamma}})_+$, and transform them with the variance function to obtain the covariate specific prior parameters:
\begin{align}
    \hat{\lambda}_k = h^{-1}(\hat{v}_k).
\end{align}
This transformation could be used to transform the prior variance estimates for the generalised additive co-data model in Equation \eqref{eq:estGAM} and for the shape-constrained co-data model in Equation \eqref{eq:estSCAM} too. Note that, however, the penalisation and constraints are applied to $\bs{\gamma}$ and not to $\bs{\lambda}$. 
So in general, for priors other than the normal prior, the variance function is not linear, such that the additivity in GAMs is on the prior variance level and not on the transformed level:
\begin{align*}
    \hat{\bs{\lambda}}=h^{-1}(\hat{\bs{v}}) = h^{-1}\left(\sum_g \hat{s}(\bs{Z}_g)\right)\neq \sum_g h^{-1}(\hat{s}(\bs{Z}_g)),
\end{align*}
nor does the transformation of a shape-constrained function, $h^{-1}(s(\bs{Z}_k))$, necessarily have the same shape as $s(\bs{Z}_k)$. 

\subsubsection{Short example for elastic net}
We may use the \proglang{R}-package \pkg{squeezy} to transform ridge penalties to elastic net penalties \citep{vanNee2021fast}, in which we use the fit from \fct{ecpc} from above.
As example, we use the elastic net parameter $\alpha=0.3$, for which we summarise the obtained elastic net penalties and regression coefficients:
\begin{CodeChunk} 
\begin{CodeInput}
R> if(!requireNamespace("squeezy")) install.packages("squeezy")
R> library("squeezy")
R> fit.EN <- squeezy(Y, X, alpha=0.3, X2=X2, Y2=Y2, lambdas=fit$penalties)
R> summary(fit.EN$lambdapApprox) #transformed elastic net penalties
\end{CodeInput}
\begin{CodeOutput}
   Min. 1st Qu.  Median    Mean 3rd Qu.    Max. 
  33.19  100.45  520.73     Inf     Inf     Inf
\end{CodeOutput}
\begin{CodeInput}
R> summary(fit.EN$betaApprox) #fitted elastic net regression coefficients
\end{CodeInput} 
\begin{CodeOutput}
      Min.    1st Qu.     Median       Mean    3rd Qu.       Max. 
-0.2677508  0.0000000  0.0000000  0.0008475  0.0000000  0.3351423
\end{CodeOutput}
\end{CodeChunk}

\section{Simulation study}\label{sec:simulation}
Estimation and prediction performance have been compared for several methods in \cite{van2021flexible}. Here, we focus on continuous co-data to exemplify the newly proposed co-data models.
First, we perform a simulation study to compare the estimates of the prior variance and prediction performance for different co-data models proposed here and the adaptive discretisation proposed in the first version of \pkg{ecpc}. 
Secondly, we perform a simulation study to compare different variable selection methods.

\subsection{Estimation and prediction performance of various co-data models}
We use the same simulation set-up as in \cite{van2021flexible} and simulate $50$ training and test data sets for some true vector of regression coefficients $\bs{\beta}^0\in\mathbb{R}^{300}$. 
Again, we consider random and informative co-data, but now continuous versions of it:
\begin{enumerate}[noitemsep,nolistsep]
\item \texttt{Random}: generate standard normal co-data $Z_k\overset{i.i.d.}{\sim}N(0,1)$ for $k=1,..,p$. 
\item \texttt{Informative}: use the true regression coefficients to inform the co-data; $Z_k=|\beta^0_k|$.
\end{enumerate}
We compare the following co-data models:
\begin{enumerate}[label=\roman*),noitemsep,topsep=0pt]
    \item \texttt{ridge}: a co-data agnostic, global ridge penalty, corresponding to the co-data intercept only model. Any co-data method should outperform this baseline method when co-data is informative, and preferably not lose much when co-data is not informative;
    \item \texttt{linear}: a linear co-data model with an intercept and one (non-)informative co-data variable;
    \item \texttt{gam}: a generalised additive co-data model using p-splines of degree $3$ and with difference penalty matrix of second order differences as suggested in \cite{eilers2021practical}. We use $20$ splines and the marginal likelihood method available in \fct{bam} from the \pkg{mgcv} package unless stated otherwise;
    \item \texttt{scam.p}: same as the generalised additive model but with shape constrained to be positive;
    \item \texttt{scam.pmi}: same as the generalised additive model but with shape constrained to be positive and monotonically increasing;
    \item \texttt{AD}: adaptive discretisation of the continuous co-data as proposed in \cite{van2021flexible}. We use a minimum of $20$ variables in the smallest groups, which leads to seven hierarchical groups.
\end{enumerate}





\begin{figure}
    \centering
    \includegraphics[width=\textwidth]{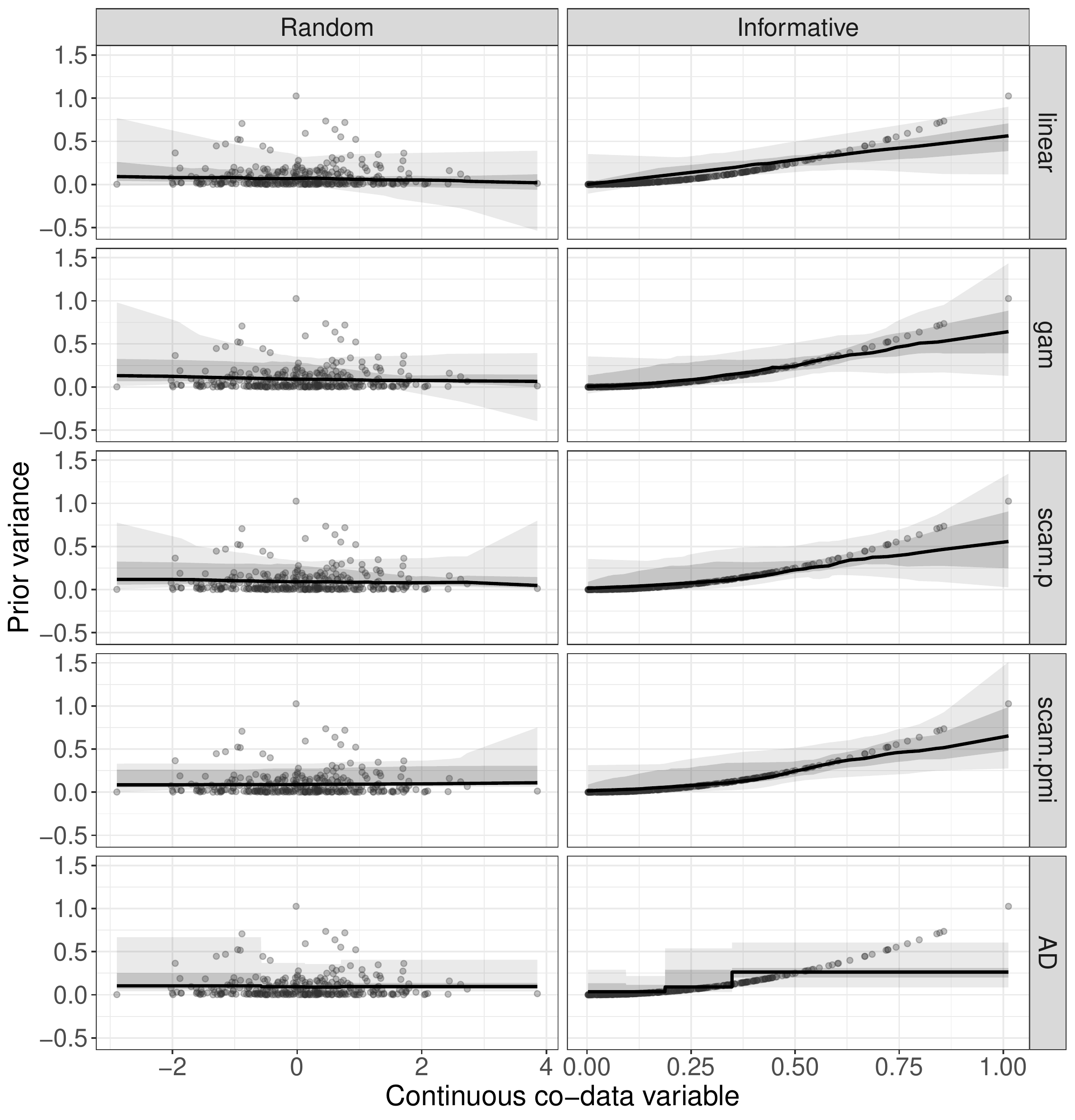}
    \caption{Simulation study based on 50 training and test sets and random co-data (left) or informative co-data (right). Estimated prior variance for various co-data models. The lines indicate the pointwise median and the inner and outer shaded bands indicate the 25-75\% and 5-95\% quantiles respectively. Points indicate the true $(\beta_k^0)^2$.}
    \label{fig:estAll}
\end{figure}
\begin{figure}
    \centering
    \includegraphics[width=\textwidth]{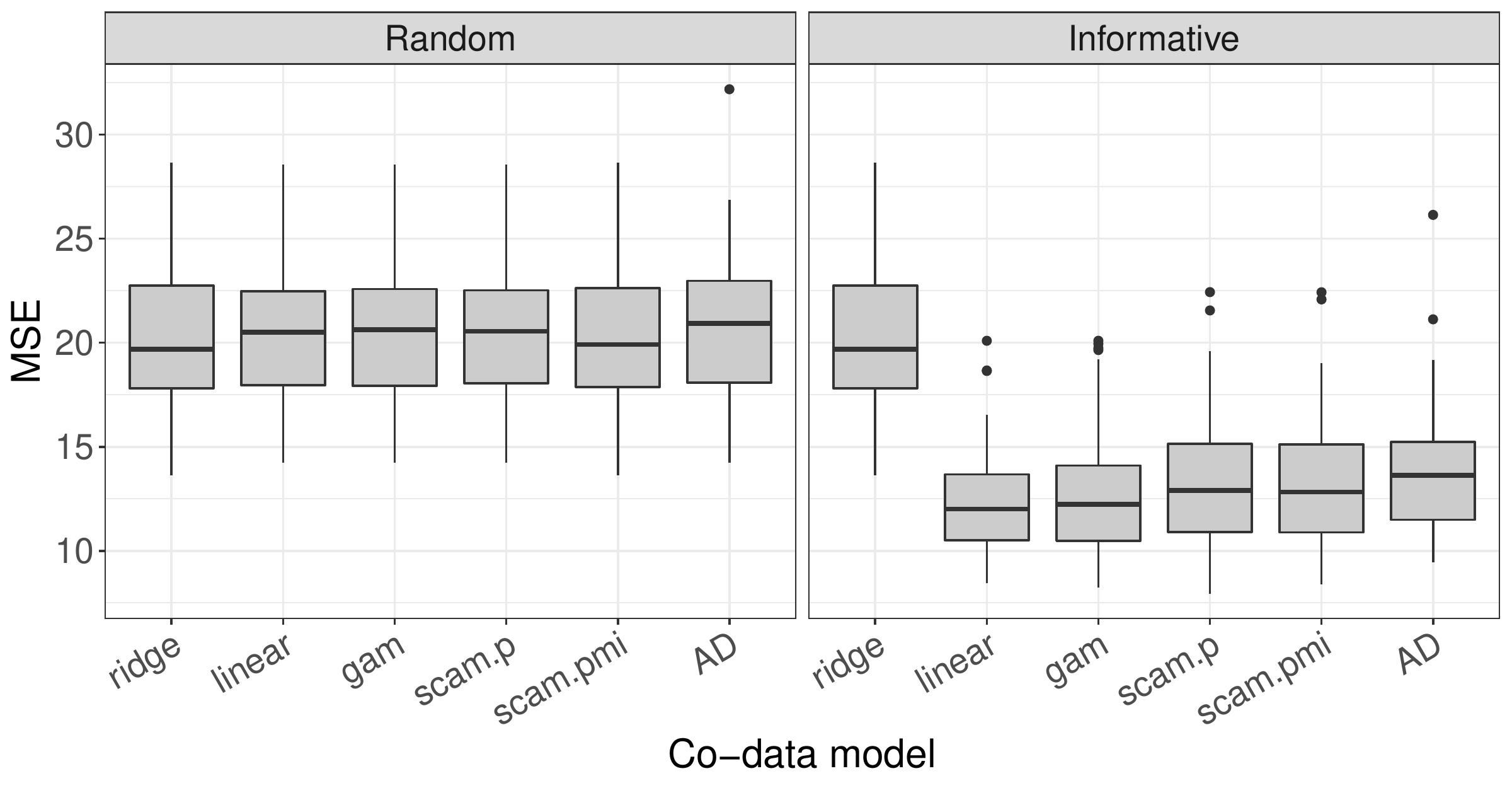}
    \caption{Simulation study based on 50 training and test sets and random co-data (left) or informative co-data (right). Boxplots of the MSE of the predictions on the test sets for various co-data models.}
    \label{fig:predAll}
\end{figure}

Figure \ref{fig:estAll} shows prior variance estimates for different co-data models with corresponding prediction performance shown in Figure \ref{fig:predAll}. 
As expected, the estimated prior variance is flat for random co-data and increasing for informative co-data, leading to prediction performance similar to and better than the co-data agnostic ordinary ridge, respectively.
The estimates of the (constrained) generalised additive co-data models are slightly more non-linear than the linear estimate, but lead to similar prediction performance. 
The variance of the estimates of the constrained generalised additive models in the random co-data reflects the effect of adding constraints, e.g. the estimates vary mostly in the positive direction for the positively constrained model. 
The linear and (constrained) generalised additive model slightly outperform the adaptive discretisation.
One advantage of the additive models using p-splines over the adaptive discretisation is that the p-splines can estimate local changes on a finer level; while the adaptive discretisation method is limited to discretisations in which each discretised group contains at least one variable (in our case, at least 20 variables per discretised group), this is not needed for p-splines, as they are penalised with a difference penalty.
To illustrate, Figure \ref{fig:estPredGAMsb} in Appendix \ref{ap:simulations} shows the generalised additive model estimates in one training data set when $G=20$ or $G=50$ splines are used and when the difference penalty is estimated with one of the methods provided in the \proglang{R}-function \fct{bam} or with the random splits as used in the first version of \pkg{ecpc}.
Except for the generalised cross-validation criterion \code{GCV.Cp}, the estimates and corresponding prediction performance seem to be robust for the number of splines (see also Figures \ref{fig:estGAMs} and \ref{fig:estPredGAMsa} in Appendix \ref{ap:simulations}).
Finally, Table \ref{tab:SimTimes} shows the average run times of the methods. The adaptive discretisation is around 3-6 times as slow as the (constrained) additive co-data models.

\begin{table}[t!]
\centering
\begin{tabular}{llll}
\textbf{Co-data model} & \textbf{bam.method} & \textbf{G} & \textbf{Time (s)} \\ \hline
linear & none & 2 & 12.7 (4.5) \\
gam & fREML & 20 & 12.3 (5.6) \\
gam & fREML & 50 & 13.0 (5.6) \\
gam & GCV.Cp & 20 & 11.6 (5.5) \\
gam & GCV.Cp & 50 & 12.1 (5.8) \\
gam & ML & 20 & 12.5 (5.6) \\
gam & ML & 50 & 12.9 (5.6) \\
gam & splits & 20 & 13.2 (5.5) \\
gam & splits & 50 & 16.5 (5.8) \\
scam.p & splits & 20 & 16.5 (11.5) \\
scam.p & splits & 50 & 19.8 (10.6) \\
scam.pmi & splits & 20 & 20.1 (17.6) \\
scam.pmi & splits & 50 & 26.2 (20.4) \\
AD & splits & 7 & 72.1 (9.8)
\end{tabular}%
\caption{Results for simulation study based on 50 training and test sets. Average run time and standard deviation for various co-data models, smoothing parameter estimation methods (as used in \fct{bam} from \pkg{mgcv} or with splits in \pkg{ecpc}) and number of co-data variables.}
\label{tab:SimTimes}
\end{table}



\subsection{Variable selection compared to other methods}
We alter the simulation set-up from above for variable selection. We now set 250 regression coefficients to $0$, leaving 50 non-zero coefficients. We scale the regression coefficients such that the L2-norm of $\bs{\beta}^0$ and the scaled, sparse $\bs{\beta}^{0,s}$ are the same.
We use the following co-data:
\begin{enumerate}[noitemsep,nolistsep]
\item \texttt{Random}: as in the simulation study above, so $Z_k^{(1)}\overset{i.i.d.}{\sim} N(0,1)$ for $k=1,..,p$. 
\item \texttt{Informative+monotone}: as in the simulation study above, but with white noise added such that the co-data is not exactly $0$ for the zero coefficients, $Z_k^{(2)}\overset{ind.}{\sim} N(|\beta^{0,s}_k|,\sigma_0^2)$, for $\sigma_0$ a tenth of the sample standard deviation of $\bs{\beta}^{0,s}$. The effect size $|\beta_k^{0,s}|^2$ is (up till some noise) a monotone, quadratic function of the co-data.
\item \texttt{Informative+convex}: similar to the \texttt{Informative} co-data but distinguishing negative from positive effects by $Z_k^{(3)}= sign(\beta^{0,s}_k)\cdot Z_k^{(2)}$. The effect size $|\beta_k|^2$ is now (up till some noise) not monotone but a convex, quadratic function of the co-data.
\end{enumerate}
We compare the following variable selection methods:
\begin{enumerate}[label=\roman*),noitemsep,topsep=0pt]
    \item \texttt{glmnet}: a co-data agnostic elastic net model fitted with \pkg{glmnet} \citep{friedman2010regularization}. 
    \item \texttt{fwelnet}: an elastic net model with continuous co-data, fitted with \pkg{fwelnet} \citep{tay2020feature}. The elastic net penalties are a fixed, exponential function of the co-data weights. 
    \item \texttt{ecpc+squeezy}: a GAM for the co-data fitted with \pkg{ecpc}, followed by a transformation of the ridge penalties to elastic net penalties by \pkg{squeezy} \citep{vanNee2021fast} for variable selection.
    \item \texttt{ecpc+postselection}: a GAM for the co-data, using the default option for posterior selection provided in the \pkg{ecpc} software.
\end{enumerate}
The first three methods have one additional tuning parameter, the elastic net parameter $\alpha\in[0,1]$, with $0$ corresponding to the full model and $1$ to the lasso model. The last method has one tuning parameter, the number of selected covariates (or equivalently, the proportion of estimated zero effects), ranging from 300 to 0 (0 to 1) for the full model to the most sparse model. In practice, one may choose one value from a range of values for the tuning parameter by comparing predictive performances and selecting the sparsest model that performs (nearly) optimal.

\begin{figure}
    \centering
    \begin{subfigure}[c]{\textwidth}
    \centering
    \includegraphics[width=\textwidth]{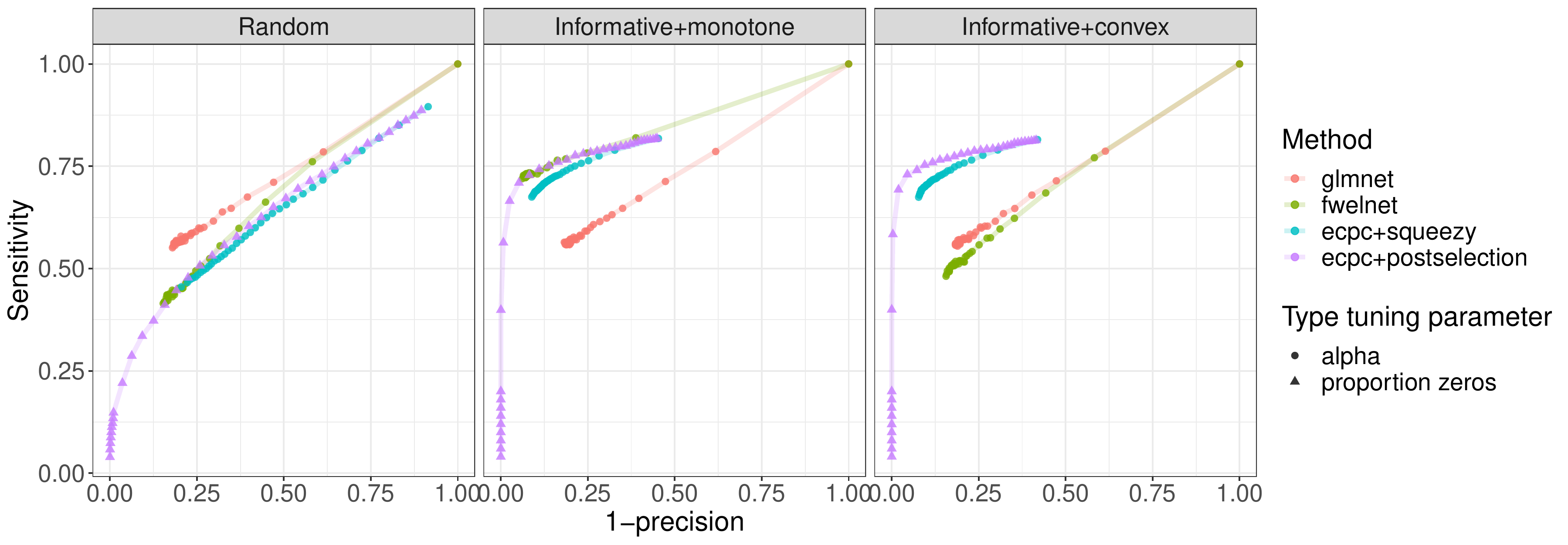}
    \caption{}
    \end{subfigure}
    \begin{subfigure}[c]{\textwidth}
    \centering
    \includegraphics[width=\textwidth]{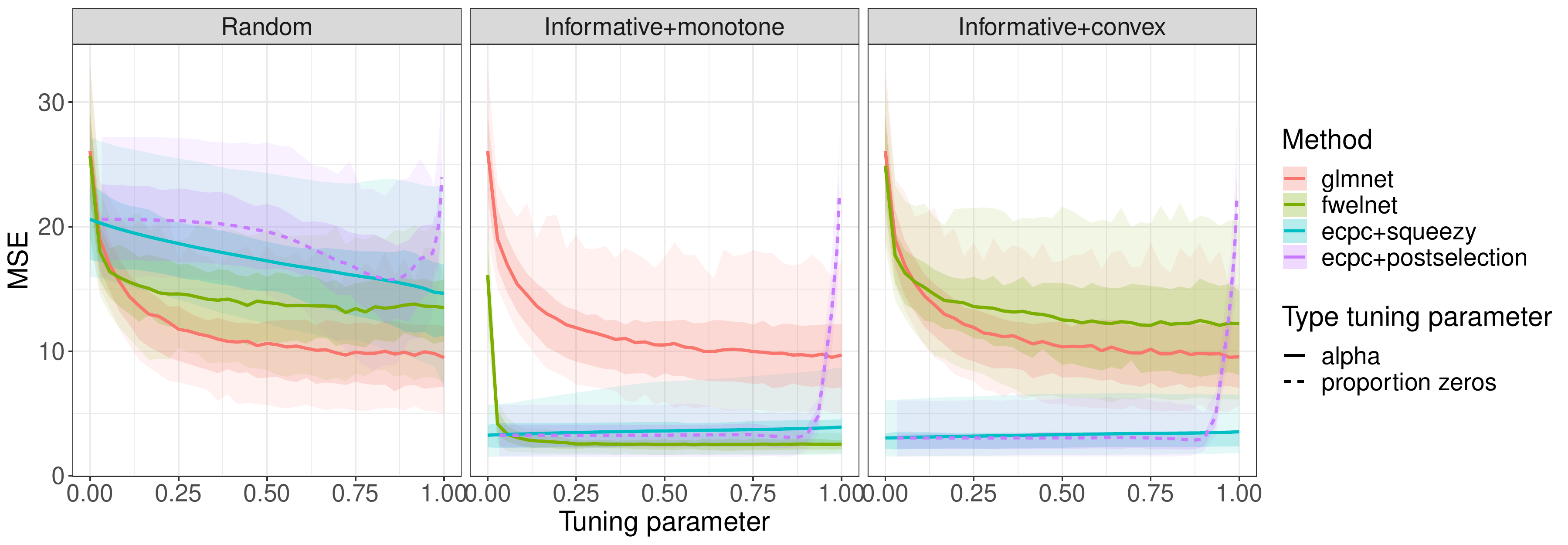}
    \caption{}
    \end{subfigure}
    \caption{Simulation study for variable selection based on 50 training and test sets for various types of co-data. a) Average sensitivity and precision for several methods and various tuning parameters; b) Mean squared error prediction performance on the test data. The lines indicate the pointwise average and the inner and outer shaded bands indicate the 25-75\% and 5-95\% quantiles respectively.}
    \label{fig:PredSparseSim}
\end{figure}

Figure \ref{fig:PredSparseSim} shows the performance of the methods in variable selection and prediction error on the test data.
Note that \texttt{ecpc+postselection} may be tuned to select sparse models up to a model that is almost empty, reaching a sensitivity and 1-precision of $0$. In contrast, the models selected by the other methods still contain more variables, even in the most sparse models for tuning parameter $\alpha=1$.
Besides, \texttt{ecpc+squeezy} and \texttt{ecpc+postselection} do not always select a full model, explaining why the sensitivities do not reach $1$. This is a result from \fct{ecpc} truncating estimated negative prior variances to $0$, deselecting some variables a priori.
In the sparse setting, the co-data agnostic \texttt{glmnet} outperforms the other methods both in terms of variable selection and prediction performance when the co-data is random. This in contrast to the dense simulation setting, in which the prediction performance of \texttt{glmnet} and \texttt{gam} were on par (Figure \ref{fig:predAll}). 
For the \texttt{informative+monotone} co-data, \texttt{fwelnet} slightly outperforms \texttt{ecpc+squeezy} and \texttt{ecpc+postselection}, all outperforming \texttt{glmnet}. 
For the \texttt{informative+convex} co-data, however, \texttt{fwelnet} is not able to flexibly adapt to the convex shape of the co-data, while the flexible GAM for the co-data in \texttt{ecpc+squeezy} and \texttt{ecpc+postselection} still adequately exploits the co-data.

\section{Analysis example}\label{sec:LNManalysis}
We demonstrate the different co-data models by applying the method to an application in classifying lymph node metastasis (coded $1$) from other types of cancer (coded $0$). 
We use the data readily available from the \proglang{R}-package \pkg{CoRF}, providing high-dimensional RNA expression training data for $p=12838$ probes and $n=133$ patients, and validation data for $n_2=97$ patients. 
First we install and load the package. Then we load the data and transform the response of the validation data set to match the format of the training data:
\begin{CodeChunk} 
\begin{CodeInput}
R> if(!requireNamespace("devtools")) install.packages("devtools")
R> library("devtools")
R> install_github("DennisBeest/CoRF") 
R> library("CoRF") 
R> data("LNM_Example")
R> RespValidationNum <- as.numeric(RespValidation)-1 
\end{CodeInput}  
\end{CodeChunk}
The data provide three different sources of co-data:
\begin{enumerate}[noitemsep,nolistsep]
\item \texttt{Signature}: a published signature of genes. Probes either match a gene in the signature or not.
\item \texttt{Correlation}: cis-correlations between RNA expression and copy number.
\item \texttt{P-values}: p-values from an external, similar study, using a different technique to measure RNA expression.
\end{enumerate}
First, we prepare the co-data. 
The first co-data source is categorical. We use the helper functions \fct{createGroupset} and \fct{createZforGroupset} to transform the vector of categories to a group set and co-data matrix:
\begin{CodeChunk}
\begin{CodeInput}
R> GroupsetSig <- createGroupset(as.factor(CoDataTrain$RoepmanGenes)) 
\end{CodeInput}
\begin{CodeOutput}
[1] "Summary of group sizes:"
    0     1 
12324   514 
\end{CodeOutput}
\begin{CodeInput}
R> Z_sig <- createZforGroupset(GroupsetSig)
\end{CodeInput}
\end{CodeChunk}

The second co-data source with correlations is continuous. We use $20$ splines to flexibly model the relation between the prior variance and the correlations.
Furthermore, we constrain the relation to be positively monotonically increasing, as we expect larger correlations to be of more importance.
The co-data spline basis matrix, difference penalty matrix and constraints are obtained by:
\begin{CodeChunk} 
\begin{CodeInput}
R> G <- 20 #number of splines
R> Z_cor <- createZforSplines(values=CoDataTrain$Corrs, G=G)
R> S1_cor <- createS(orderPen=2, G=G) 
R> Con_cor <- createCon(G=G, shape="positive+monotone.i")
\end{CodeInput}
\end{CodeChunk}

We prepare the co-data with p-values similarly to the correlations, but constrain the relation to be positively monotonically decreasing, as we expect smaller p-values to be of more importance. We set the p-value of two variables that have missing p-values to the maximum observed p-value and compute the co-data (related) matrices:
\begin{CodeChunk} 
\begin{CodeInput}
R> CoDataTrain$pvalsVUmc[is.na(CoDataTrain$pvalsVUmc)] <-
+   max(CoDataTrain$pvalsVUmc, na.rm=TRUE)
R> Z_pvals <- createZforSplines(values=CoDataTrain$pvalsVUmc, G=G)
R> S1_pvals <- createS(G=G) 
R> Con_pvals <- createCon(G=G, shape="positive+monotone.d")
\end{CodeInput}  
\end{CodeChunk}
As the last step of the preparation of the co-data, we save the continous co-data variables in a separate list for the \fct{plot} function that we use below, and concatenate the co-data matrices in a list:
\begin{CodeChunk} 
\begin{CodeInput}
R> values <- list("Signature" = NULL, 
+                 "Correlation" = CoDataTrain$Corrs,
+                 "P-values" = CoDataTrain$pvalsVUmc)
R> Z_all <- list("Signature" = Z_sig, 
+                "Correlation" = Z_cor,
+                "P-values" = Z_pvals)
\end{CodeInput}  
\end{CodeChunk}
Then we fit the model and select variables a posteriori, with a range of the maximum number of variables that should be selected defined in the input argument \code{maxsel}: 
\begin{CodeChunk} 
\begin{CodeInput}
R> set.seed(3)
R> maxSel <- c(2:10,10*(2:10)) #maximum posterior selected variables
R> Res<-ecpc(Y=RespTrain, X=TrainData, Z=Z_all, 
+            paraPen = list(Z2=list(S1=S1_cor), Z3=list(S1=S1_pvals)),
+            paraCon = list(Z2=Con_cor, Z3=Con_pvals),
+            Y2=RespValidationNum, X2=ValidationData, 
+            maxsel=maxSel)
\end{CodeInput}
\begin{CodeOutput}
[1] "Estimate global tau^2 (equiv. global ridge penalty lambda)"
[1] "Co-data matrix 1: estimate weights, hypershrinkage type: none"
[1] "Co-data matrix 2: estimate hyperlambda for ridge+constraints hypershrinkage"
[1] "Estimate weights of co-data source 2"
[1] "Co-data matrix 3: estimate hyperlambda for ridge+constraints hypershrinkage"
[1] "Estimate weights of co-data source 3"
[1] "Estimate co-data source weights"
[1] "Estimate regression coefficients"
[1] "Sparsify model with posterior selection"
\end{CodeOutput}
\end{CodeChunk}
We plot the contributions from each co-data source, shown in Figure \ref{fig:plot_example}:
\begin{CodeChunk} 
\begin{CodeInput}
R> plot(Res, show="priorweights", Z=Z_all, values=values)
\end{CodeInput}  
\begin{figure}[h]
    \centering
    \includegraphics[width=\textwidth]{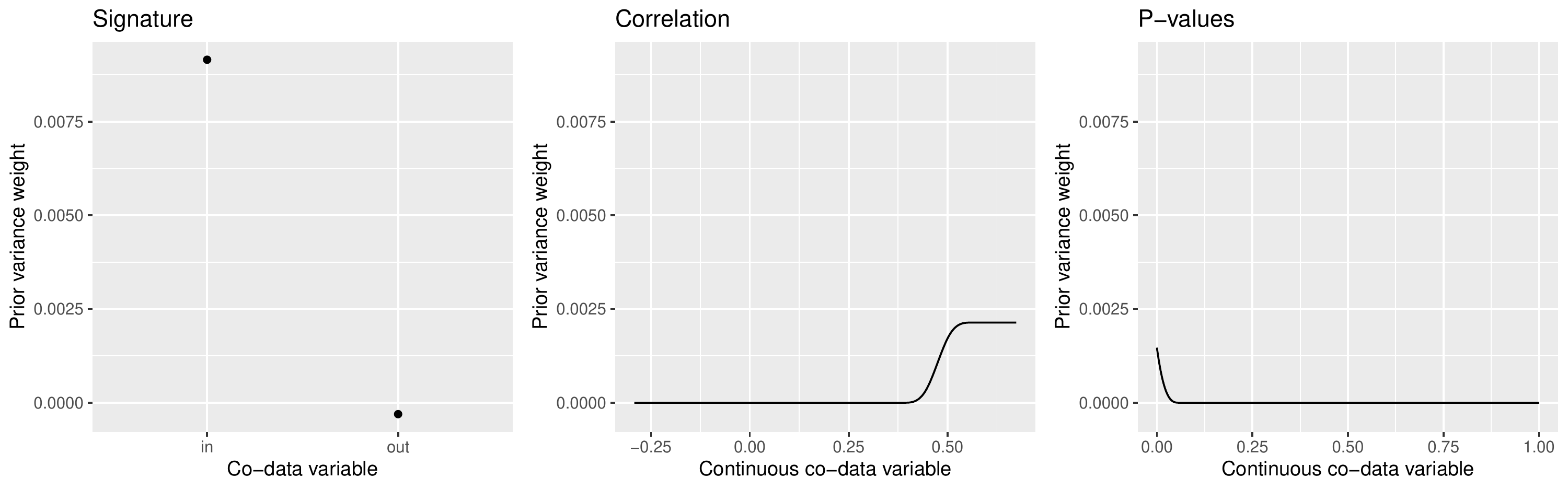}
    \caption{Data analysis example. Figure produced by the \fct{plot}-method applied on the \fct{ecpc} fit.}
    \label{fig:plot_example}
\end{figure}
\end{CodeChunk}
The predicted values for the validation data are given in \code{Res\$Ypred} as \code{X2} was provided to \fct{ecpc}. Alternatively, the predictions may be retrieved with the method \fct{predict}:
\begin{CodeChunk}
\begin{CodeInput}
R> Ypred <- predict(Res, X2=ValidationData)
\end{CodeInput}
\end{CodeChunk}
The posterior selected variables are given in \code{Res\$betaPost} as \code{maxsel} was provided to \fct{ecpc}. Alternatively, the same posterior selection method may be performed with the function \fct{postSelect} on the fitted \class{ecpc}-object \code{Res}:
\begin{CodeChunk}
\begin{CodeInput}
R> sparseModels <- postSelect(Res, X=TrainData, Y=RespTrain, maxsel=maxSel)
\end{CodeInput}
\end{CodeChunk}
A second approach for variable selection is to use \fct{squeezy} to transform the ridge penalties to elastic net penalties, with elastic net tuning parameter $\alpha$.
As mentioned above, in practice one may try a range of tuning parameters to choose the sparsest model with close to optimal performance.
For this example, we include the lasso model ($\alpha=1$):
\begin{CodeChunk}
\begin{CodeInput}
R>   sparseModel2 <- squeezy(Y=RespTrain, X=TrainData, alpha=1, 
+                          lambdas=Res$penalties,
+                          X2=ValidationData, Y2=RespValidationNum)
\end{CodeInput}
\end{CodeChunk}

\begin{figure}
    \centering
    \begin{subfigure}[c]{\textwidth}
    \centering
    \includegraphics[width=\textwidth]{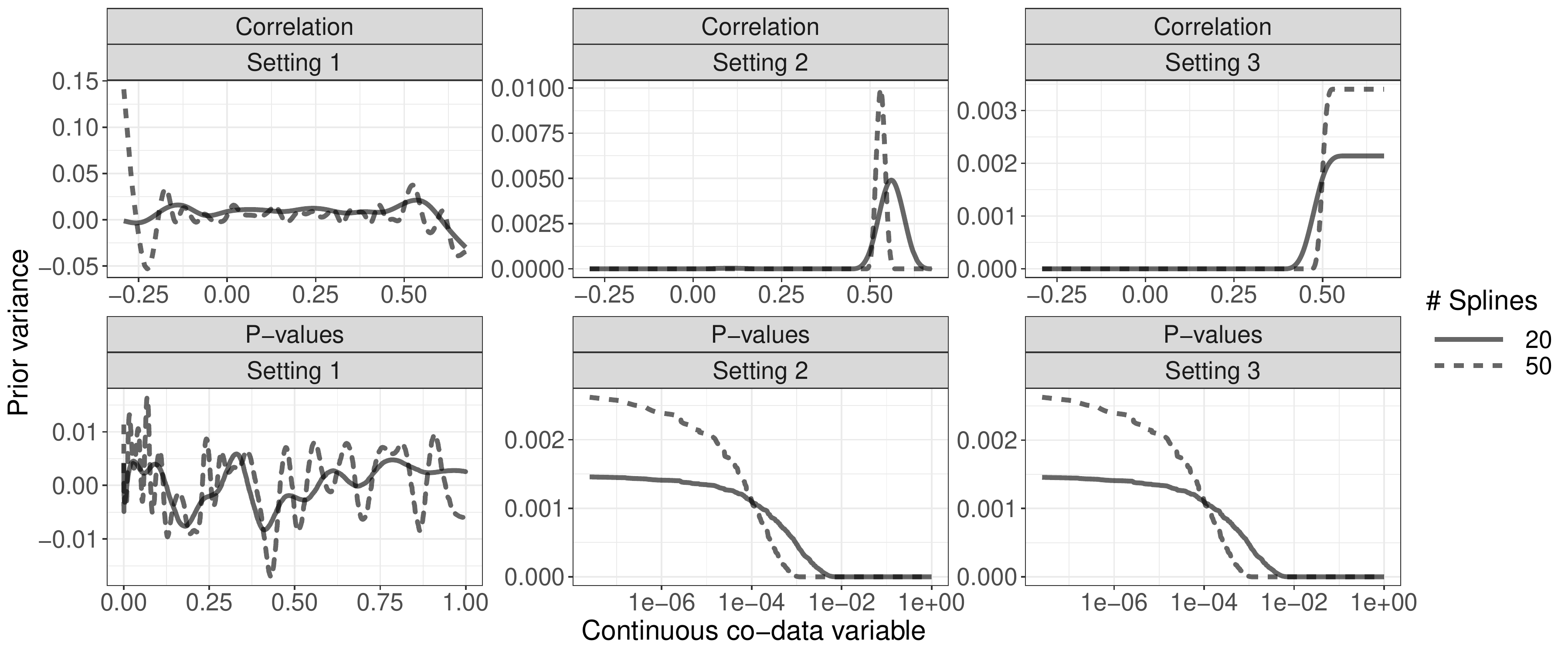}
    \caption{}
    \end{subfigure}
    \begin{subfigure}[c]{\textwidth}
    \centering
    \includegraphics[width=\textwidth]{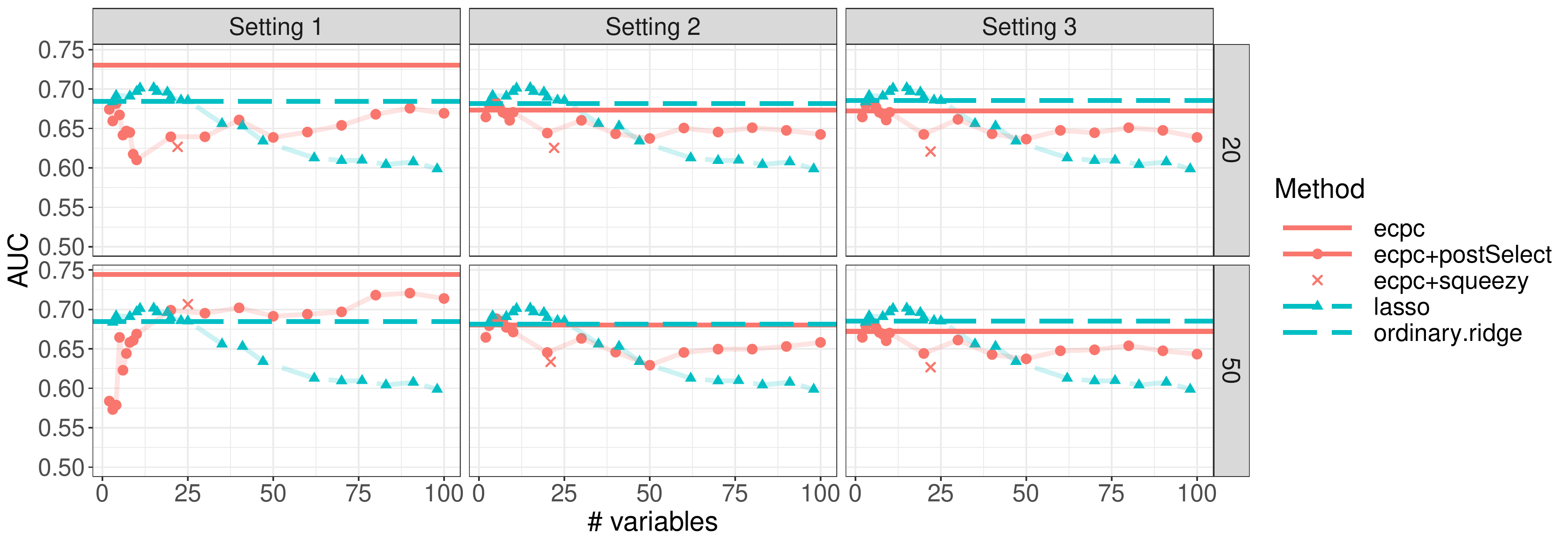}
    \caption{}
    \end{subfigure}
    \caption{Data analysis example: a) Estimated prior variance contributions of each co-data source, before multiplying with the co-data specific weight. Note that the p-values are shown on the log-scale in Settings 2 and 3, to clearly show the non-zero peaks at the smallest p-values; b) corresponding prediction performance on the validation set for $20$ or $50$ spline basis functions. The settings correspond to different co-data models: 1) no constrains; 2) positive constrained shape; 3) positive and monotonically constrained shape.}
    \label{fig:LNMExample}
\end{figure}

Instead of fitting monotone and positively constrained functions for the correlation and p-values co-data, one could consider other co-data models. 
Figure \ref{fig:LNMExample} shows the results for three different settings: 1) a GAM, i.e. without constraints; 2) a SCAM with positivity constraints; 3) a SCAM with positivity and monotonicity constraints, as in the exemplified code above.
The results include the dense models obtained with \fct{ecpc} and a co-data agnostic ordinary ridge model, and sparse models for a range of posterior selected variables obtained with \fct{postSelect}, the lasso model obtained with transformed penalties from \fct{squeezy} and a co-data agnostic lasso model.
The estimated prior variance contribution for the correlation co-data shows large deviations near the boundaries, which increase when 50 instead of 20 splines are used (Setting 1). 
While p-splines have no boundary effects in regular regression models \citep{psplines1996Eilers}, these effects may be the result from the signal regression nature of the model used in Equation \eqref{eq:estGAM}. To dampen the boundary effects, it may be beneficial to transform the co-data values such that the values spread out more evenly, e.g. using the empirical cumulative distribution function.
Fitting a co-data model with positive (and monotone) constraints (Setting 2 and 3) results in smoother functions than when it is fit without constraints.
While adding the constraints stabilises posterior selection for highly sparse models, it generally does not benefit the prediction performance when compared to the GAM co-data model.
The GAM co-data model results in the best prediction performance among sparse models, though, in practice, the simpler lasso model may be preferred as it shows competitive performance.
The overall best prediction performance on the test data is retrieved by the full, dense model when the (unconstrained) generalised additive co-data model is used with $50$ splines.

\section{Conclusion}\label{sec:conclusion}
We presented an extension to the \proglang{R}-package \pkg{ecpc} that accommodates linear co-data models, generalised additive co-data models and shape constrained additive co-data models for the purpose of high-dimensional prediction and variable selection. 
These co-data models are particularly useful for continuous co-data, for which an adaptive discretisation was available in the first version. The newly proposed co-data models are shown to run faster and lead to slightly better prediction performance when compared to the first version in a simulation study.
Moreover, the estimated variable-specific ridge penalties may be transformed to elastic net penalties with the \proglang{R}-package \pkg{squeezy} to allow for variable selection. We showed in a simulation study that this approach and the previously proposed posterior selection approach lead to similar performance, outperforming other methods when the effect sizes are (non-exponentially) related to the co-data.
We have provided several short examples and one analysis example to a cancer genomics application to demonstrate the code. 
Stand-alone \proglang{R}-scripts and other code files used for the simulations and examples may be found on \url{https://github.com/Mirrelijn/ecpc}. 

\section*{Acknowledgements}
The first author is supported by ZonMw TOP grant COMPUTE CANCER (40-
00812-98-16012).
The authors would like to thank Soufiane Mourragui (Netherlands Cancer Insitute) for the many worthwhile discussions.


\bibliography{BIB}


\newpage

\begin{appendix}
\setcounter{section}{0}
\setcounter{table}{0}
\renewcommand{\thetable}{\thesection\arabic{table}}%
\setcounter{figure}{0}
\renewcommand{\thefigure}{\thesection\arabic{figure}}%
\setcounter{equation}{0}
\renewcommand{\theequation}{\thesection\arabic{equation}}

\section{Additional figures to simulation study}\label{ap:simulations}
\begin{figure}
    \centering
        \begin{subfigure}[c]{\textwidth}
    \centering
    \includegraphics[width=\textwidth]{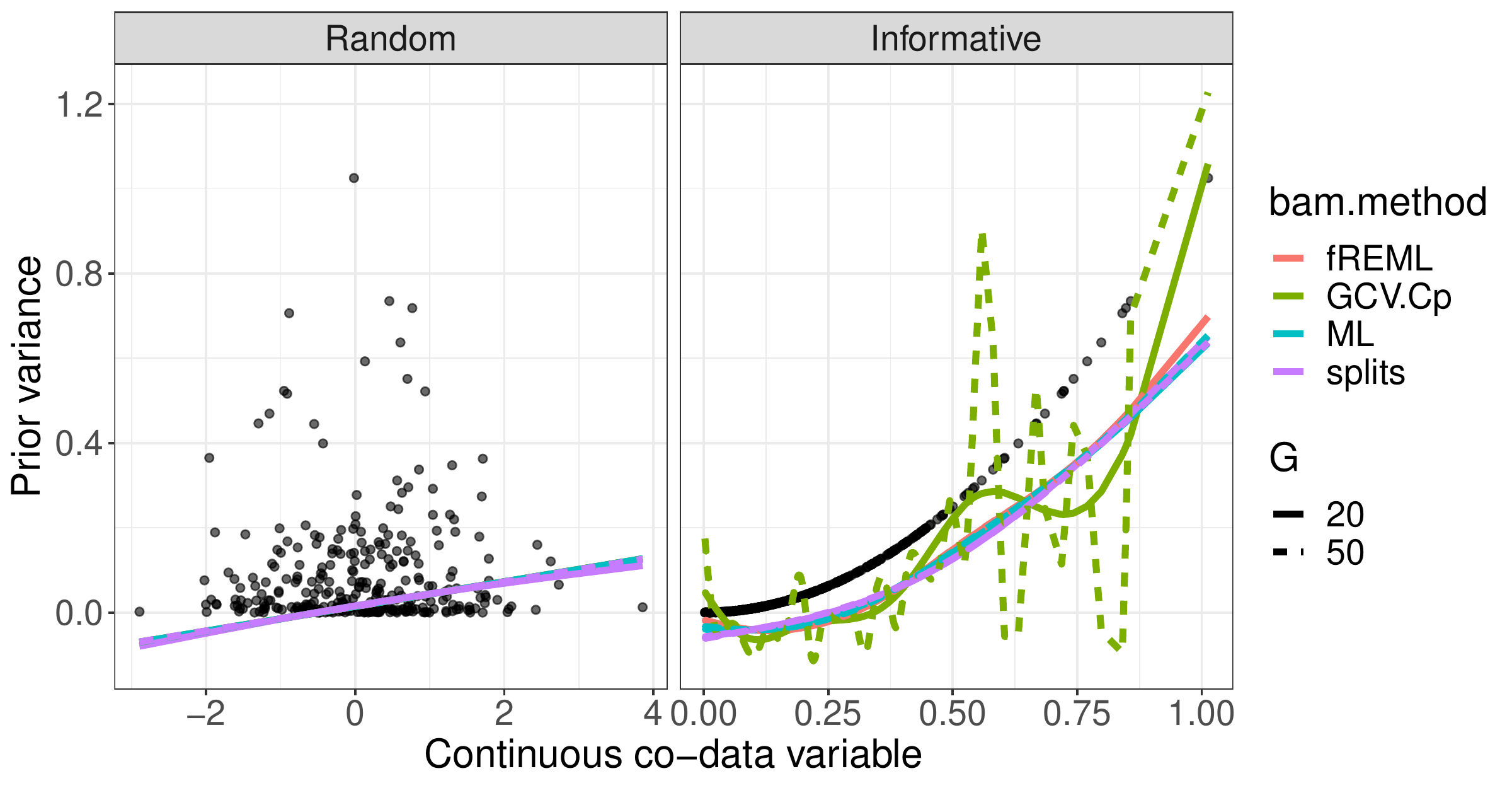}
    \caption{}
    \label{fig:estPredGAMsb}
    \end{subfigure}
    \begin{subfigure}[c]{\textwidth}
    \centering
    \includegraphics[width=\textwidth]{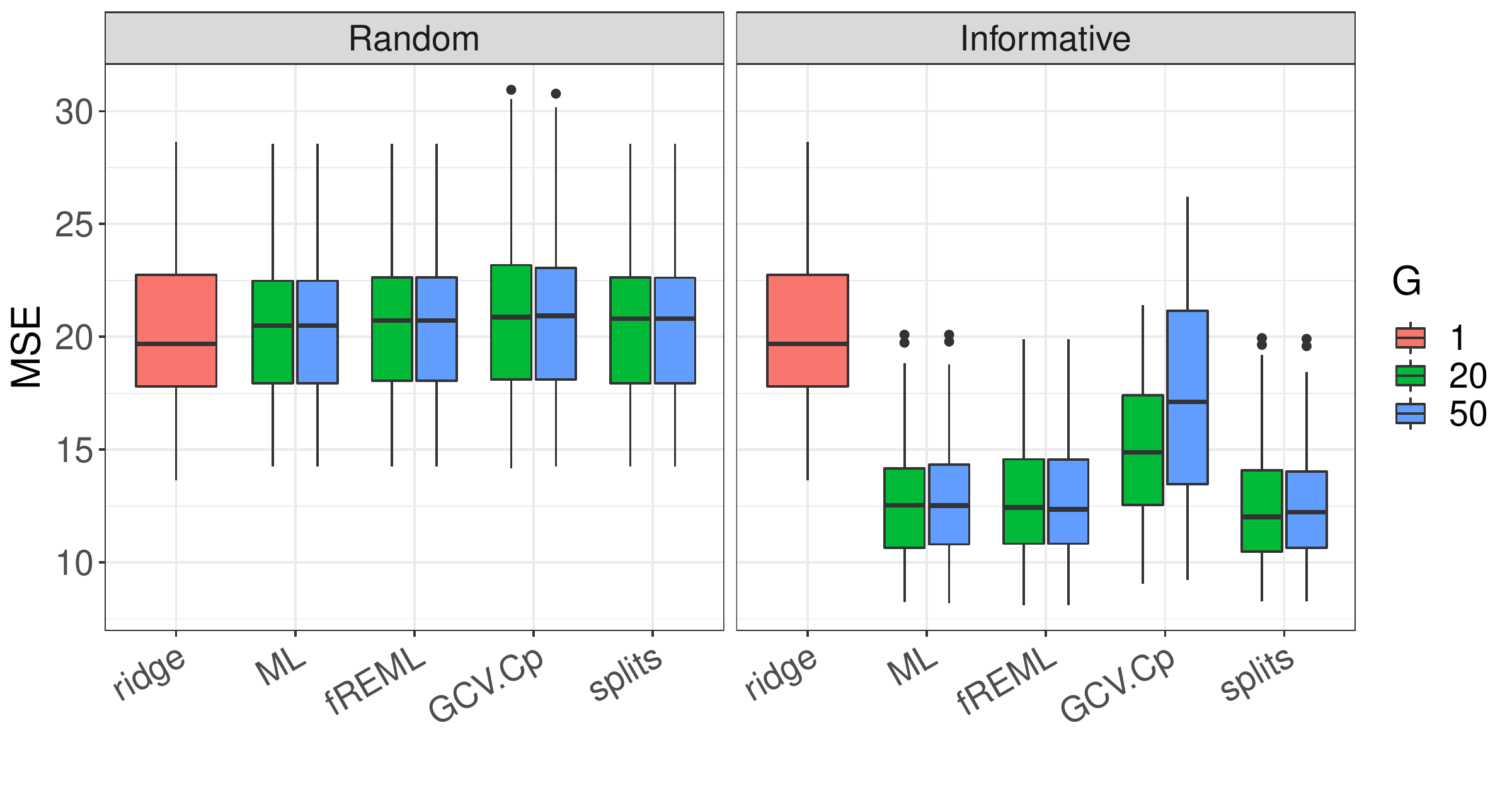}
    \caption{}
    \label{fig:estPredGAMsa}
    \end{subfigure}
    \caption{Simulation study based on 50 training and test sets and random co-data (left) or informative co-data (right). a) Example of estimated prior variance for various smoothing parameter estimation methods in one training data set; b) boxplots of the MSE of the predictions on the test sets for the ordinary ridge model ($G=1$ co-data intercept variable) and for a generalised additive co-data model using various smoothing parameter estimation methods and $G=20$ or $50$ splines.}
    \label{fig:estPredGAMs}
\end{figure}

\begin{figure}
    \centering
    \includegraphics[width=\textwidth]{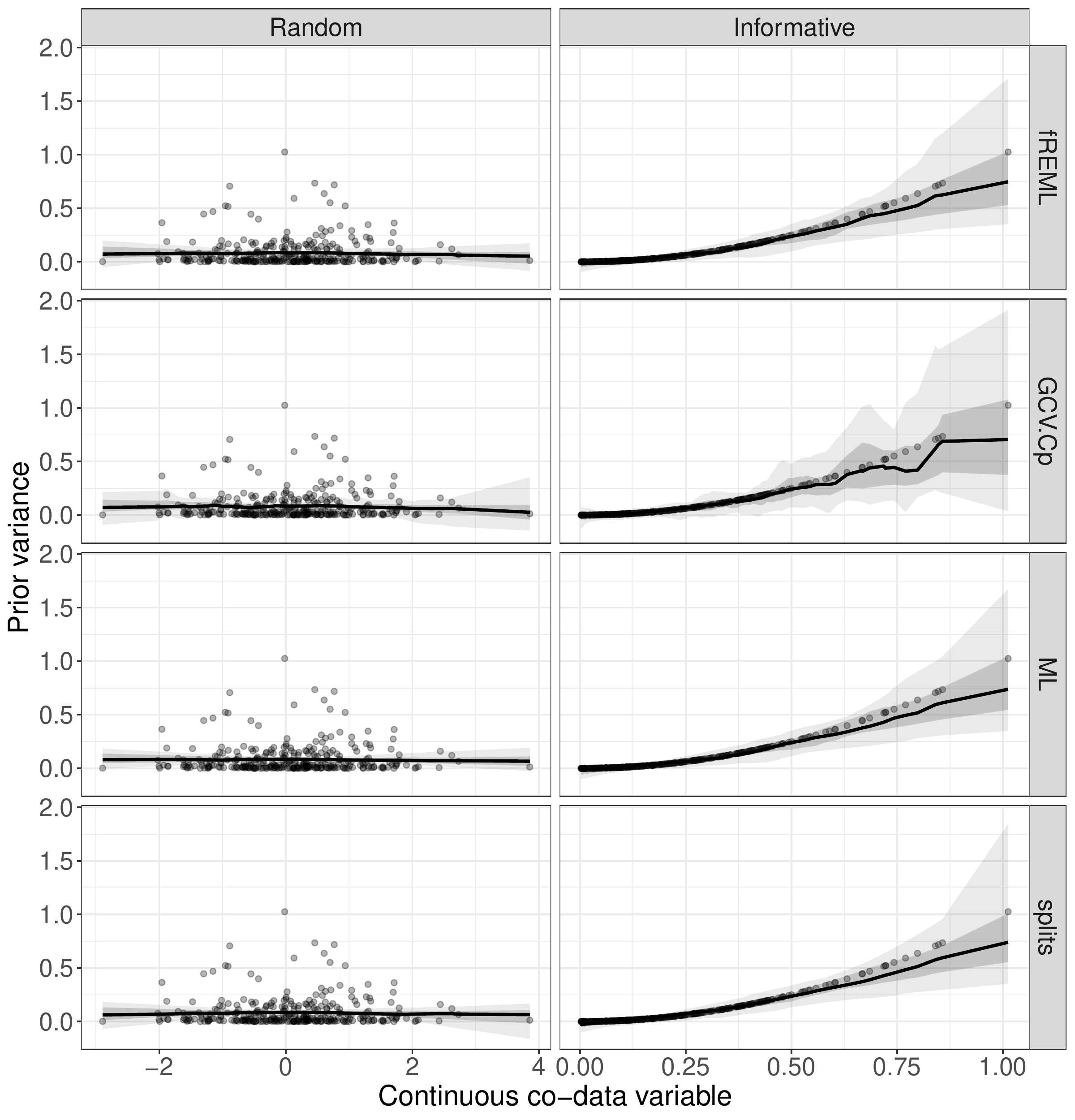}
    \caption{Simulation study based on 50 training and test sets and random co-data (left) or informative co-data (right). Estimated prior variance for the generalised additive co-data model using various smoothing parameter estimation methods. The lines indicate the pointwise median and the inner and outer shaded bands indicate the 25-75\% and 5-95\% quantiles respectively. Points indicate the true $(\beta_k^0)^2$.}
    \label{fig:estGAMs}
\end{figure}

Figures \ref{fig:estPredGAMs} and \ref{fig:estGAMs} show the results for generalised additive co-data models when different smoothing parameter methods are used.

\end{appendix}


\end{document}